\documentclass[letterpaper,12pt]{article}

\usepackage[T1]{fontenc}

\usepackage{geometry}
\usepackage{setspace}
\usepackage[normalem]{ulem}
\usepackage{textcomp}

\usepackage[style=chem-acs,backend=biber,articletitle=true]{biblatex}
\usepackage{graphicx,color}
\usepackage{float}
\newfloat{scheme}{htbp}{los}
\floatname{scheme}{Scheme}
\floatname{chart}{Chart}
\newfloat{graph}{htbp}{loh}

\usepackage{chemformula} 
\usepackage[version = 4]{mhchem} 
\usepackage{amsmath}
\usepackage{bm}
\usepackage{dcolumn}
\usepackage{ifthen}
\usepackage{multirow}
\usepackage{threeparttable}
\usepackage{siunitx}

\usepackage{hyperref}
\hypersetup{
    pdfnewwindow=true, 
    linkcolor=blue, 
    colorlinks=true, 
    anchorcolor=blue, 
    citecolor=blue, 
    filecolor=blue,  
    menucolor=blue, 
    urlcolor=blue,
}

\usepackage{authblk}
\author[1,*]{Affan Safeer}
\author[1]{Oktay G\"{u}lery\"{u}z} 
\author[1]{Guangyao Miao} 
\author[1]{Wouter Jolie}
\author[2]{Keith McKenna}
\author[1]{Thomas Michely} 
\author[1]{Jeison Fischer}
\affil[1]{II. Physikalisches Institut, Universit\"{a}t zu K\"{o}ln, Z\"{u}lpicher Stra\ss e 77, 50937 K\"{o}ln, Germany}
\affil[2]{School of Physics, Engineering and Technology, University of York, Heslington
York YO10 5DD, U.K.}

\title{Substrate manipulation of small polarons in single-layer MnBr$_2$}
\date{*Email: safeer@ph2.uni-koeln.de}

\begin{document}

\maketitle

\begin{abstract}
  Single-layer transition metal dihalides grown on conducting substrates were shown to host stable polarons. Here, we investigate small polarons in a single layer of the insulator MnBr$_2$, which was grown by molecular beam epitaxy on three different substrates: graphene on Ir(110), graphene on Ir(111), and Au(111). 
  Polarons on these substrates can be created, converted, and moved by the STM tip when a tunneling current flows at a proper bias voltage. Four polaron species are observed in MnBr$_2$ grown on graphene Ir(110), three of which resemble the species previously observed for CoCl$_2$ on graphite, whereas MnBr$_2$ grown on the other substrates exhibits only two of these four species. Changing the substrate from graphene on Ir to Au(111) increases polaron number density by an order of magnitude. These large differences demonstrate that the choice of conducting substrate plays a decisive role in determining polaron phenomenology in substrate-supported single-layer insulators. Two unique observations are reported for polarons in MnBr$_2$ on graphene supported by Ir(110): First, mobile polarons are guided by the periodic potential originating from the super-moiré of MnBr$_2$ with graphene and Ir(110); second, polarons remain stable up to 300\,K.

\end{abstract}

\section{Introduction}

An excess charge carrier (electron or hole) introduced into a polarizable material can couple to lattice vibrations and distort the surrounding ions, creating a local lattice distortion bound to the carrier. The resulting composite quasiparticle (a charge dressed by lattice distortions) is called a polaron. Depending on the spatial extent of electron-phonon coupling, one commonly distinguishes small or Holstein polarons, where the distortion is confined to one or a few lattice sites, and large or Fröhlich polarons, where both the lattice distortion and charge carrier wave function extend over many unit cells \cite{franchini2021}. Polarons can strongly modify charge transport properties by increasing effective mass or promoting thermal-activated hopping, and can also affect other properties such as chemical reactivity (see review ref.~\cite{franchini2021}).
Correspondingly, polarons are central to the functional behavior and technological use of many materials, for example, transition metal oxides \cite{Rettie16,Reticcioli19b,Tanner22,Cheng23}, where charge transport is often governed by small polarons and halide perovskites \cite{Miyata17,Ghosh20}, where large polarons are frequently invoked. 

Although polarons in solids can be detected with a multitude of methods, the ability of scanning tunneling microscopy (STM) and - spectroscopy (STS) to visualize the charge distribution at a surface has made a substantial impact on the development of the field. Considering the paradigmatic example of small polarons in rutile TiO$_2$, excess charge carriers are introduced into the unoccupied 3$d$ states of Ti atoms next to surface oxygen vacancies created, e.g., by vacuum annealing. Using STM, STS, and ab initio calculations, the resulting charge distribution was measured \cite{Minato09} and unambiguously identified as due to small polarons \cite{Setvin14}. Since then, local probe experiments have been instrumental in improving our understanding of small polarons \cite{Yim16, Reticcioli17, Reticcioli19, Birschitzky24, Yim24, Redondo24, Skreekumar25}.  

Evidently, 2D materials with their reduced screening are also candidates for polaron formation, although no evidence for polarons in 2D materials was published until 2018. A first report on a polaron in a 2D material was of interface polarons between hexagonal boron nitride and graphene, detected by angle-resolved photoemission \cite{Chen18}. Using the same method, polarons were reported in doped MoS$_2$ \cite{Kang18}, but this interpretation was convincingly questioned \cite{Garcia19,Sio23}. Recently, based on STS and STM, the existence of polarons in doped single-layer MoS$_2$ was proposed \cite{VanEfferen25}.  

A breakthrough occurred and a new paradigm evolved, when two back-to-back studies demonstrated the creation, annihilation, and manipulation of small polarons by an STM tip in epitaxially grown single-layer CoCl$_2$ \cite{cai2023, liu2023}. These findings were later extended to other members of the family of transition metal dihalides, including FeCl$_2$ \cite{cai2023}, CrI$_2$ \cite{Liang2025}, NiI$_2$ \cite{miao2025}, and CrBr$_3$ \cite{Cai2025}. 
Complementary ab initio calculations were stimulated by the observed manifold of polarons, their stability, and the ease with which they can be manipulated with the STM tip \cite{cai2023,liu2023,Yao2024, Cai2025}. In addition to the small polaron interpretation for CoCl$_2$, an alternative scenario of solitons due to charge dipoles has also been proposed \cite{Hao24,Song25}. 

Here, we investigate polarons in single-layer MnBr$_2$, a transition metal dihalide that has not yet been explored from this perspective. Mn has the lowest electronegativity among the transition metals, enhancing ionic interactions with halides and thus ionic polarizability, a key factor for polaron formation \cite{Sio23}. The half-filled 3d$^5$ configuration of Mn diminishes electron hopping, leading to a flat conduction band with d-character \cite{Jesus25}, a favorable condition for electron localization. Lastly, although electron-phonon coupling in MnBr$_2$ has not been investigated, its phonons are soft, with maximal energies on the order of 20\,meV \cite{Jesus25,Luo20}; i.e., comparatively little energy is involved in the deformation of the MnBr$_2$ lattice. 

Although our results show some similarities to those for polarons in single layers of CoCl$_2$ on graphite, we uncover a number of new aspects. The emergence of a polaron species not reported so far, the strong interaction of polarons with a super-moiré resulting from the interaction of MnBr$_2$ with its substrate, and an impressive thermal stability of polarons. Most importantly, a hitherto somewhat overlooked strong dependence of the polaron phenomenology on the substrate is identified. This substrate dependence has profound implications for understanding polaron formation. 

Single-layer MnBr$_2$ has been investigated theoretically using density functional theory (DFT) \cite{Kulish17, Botana19, Li2020, Luo20, Bo2024, Jesus25}. Mn ions with $\approx 5 \mu_\mathrm{B}$ magnetic moments are predicted in most studies to be antiferromagnetically ordered at zero temperature. However, experimentally bulk MnBr$_2$ is found to be paramagnetic above 2.35\,K \cite{Wollan58, farge1976, Iio1990, Sato1994}. In our experiments, no magnetic ordering could be detected down to 1.7\,K using spin-polarized STM \cite{Safeer2026Cr2S3, Bagchi2024, Knispel2025}. At 4.2\,K, which is the measurement temperature for the data presented here,  MnBr$_2$ is expected to be paramagnetic.

In a previous experimental study by some of us \cite{safeer25b}, single-layer MnBr$_2$ grown on Gr/Ir(110) and Gr/Ir(111) displayed the same structure as a layer in the van der Waals bulk material, that is, a hexagonal CdI$_2$-type lattice in which six Br$^-$ ions sandwich the Mn$^{2+}$ ions and coordinate them in an edge-sharing arrangement. The experimental lattice constant of 3.90\,Å closely matches the bulk lattice parameter of 3.873\,Å \cite{Ronda87}, confirming the structural integrity of MnBr$_2$ in the single-layer limit. Single-layer MnBr$_2$ is an insulator with an experimental band gap of 4.4\,eV \cite{safeer25b} on Gr/Ir(110), slightly larger than the previous DFT predictions of 4.21\,eV \cite{Luo20} or 3.95\,eV \cite{Jesus25} for the freestanding single-layer. Of specific relevance for the present work is the fact that MnBr$_2$ on Gr/Ir(110) forms a complex super-moiré \cite{safeer25b}. The super-moiré is visible only below the conduction band edge (CBE) of MnBr$_2$ located at 1.9\,eV. A key finding was that the dominant Fourier component of the MnBr$_2$/Gr/Ir(110) super-moiré is due to a virtual moiré between MnBr$_2$ and Ir(110). This Fourier component gives rise to a stripe pattern with a wavelength $\lambda_{sm} \approx 2.7$\,nm, which correspondingly dominates the real-space super-moiré pattern.

\section{Results}

\subsection{Polaron phenomenology and interaction with the super-moiré}

\begin{figure}[ht!]
\includegraphics[width=0.5\textwidth]{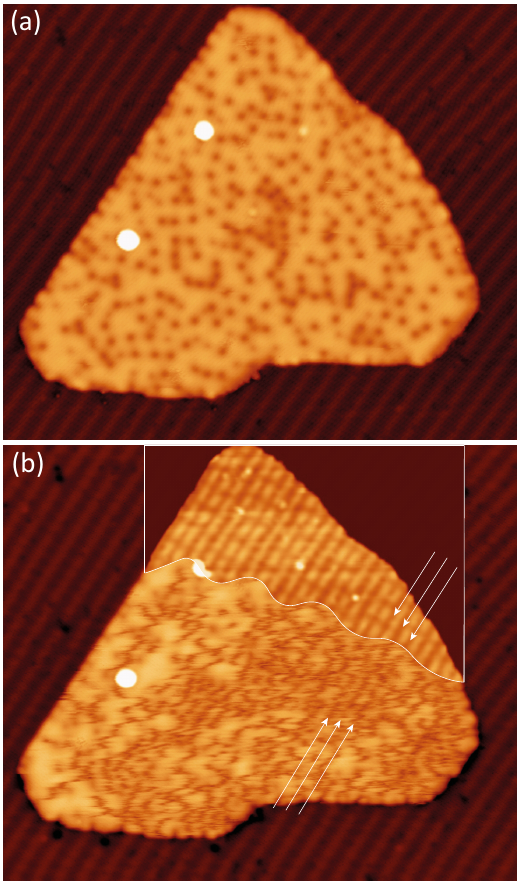}
\caption{Polaron mobility. (a) Single-layer MnBr$_2$ island on Gr/Ir(110) imaged with $V_\mathrm{b} = 2.5$\,V, $I_\mathrm{t} = 20$\,pA. At this $V_\mathrm{b}$, polarons are immobile. (b) Same island as in (a), but $V_\mathrm{b} = 3$\,V, $I_\mathrm{t} = 20$\,pA. Polarons fluctuate in position under the influence of the STM tip. The inset, bounded by a white line, displays the same island as in (a), but imaged with $V_\mathrm{b} = -1$\,V, $I_\mathrm{t} = 20$\,pA. At this voltage, the MnBr$_2$/Gr/Ir(110) super-moiré is visible.  STM topographs are taken at 4.2\,K. Images sizes: 100\,nm $\times$ 85\,nm.}
\label{Fig_Polaron_presence}
\end{figure}

Figure~\ref{Fig_Polaron_presence}(a) displays a low-temperature STM topograph of a single-layer MnBr$_2$ island acquired at $V_\mathrm{b} = 2.5$\,V. A high density of depressions is visible with a diameter of the order of $\approx 2$\,nm at half depth and a maximum apparent depth of $\approx 2$\,\AA. 
Upon increasing $V_\mathrm{b}$ to 3.0\,V, most dark spots transform into fuzzy features. These dark fuzzy features follow, not perfectly, but are well recognizable, the stripes of the MnBr$_2$/Gr/Ir(110) super-moiré with a wavelength of $\lambda_{sm} \approx 2.7$\,nm  \cite{safeer25b}. This becomes obvious through the inset of Figure~\ref{Fig_Polaron_presence}(b), which is taken at $V_\mathrm{b} = -1.0$\,V, a voltage where, for this magnification, only the dominant Fourier component of the super-moiré is visible \cite{safeer25b}. Two sets of white arrows emphasize the 2.7\,nm periodicity of the super-moiré. In the surroundings of the island, the Gr/Ir(110) moiré is recognized as described in detail in ref. \cite{Kraus22}. It has a distinctly larger wavelength $\lambda_{m}$ (typically $\approx 3.3$\,nm) than $\lambda_{sm}$.

The dark spots observed in the single-layer MnBr$_2$ island of Figure~\ref{Fig_Polaron_presence}(a) are attributed to polarons in MnBr$_2$, as discussed below. The fuzzy stripe-like features arise from polaron mobility driven by interaction with the STM tip. The fuzzy polarons feature partly align with the stripes of the super-moiré. This alignment indicates an interaction of the polarons with the MnBr$_2$/Gr/Ir(110) super-moiré. 

\begin{figure}[ht!]
\includegraphics[width=0.95\textwidth]{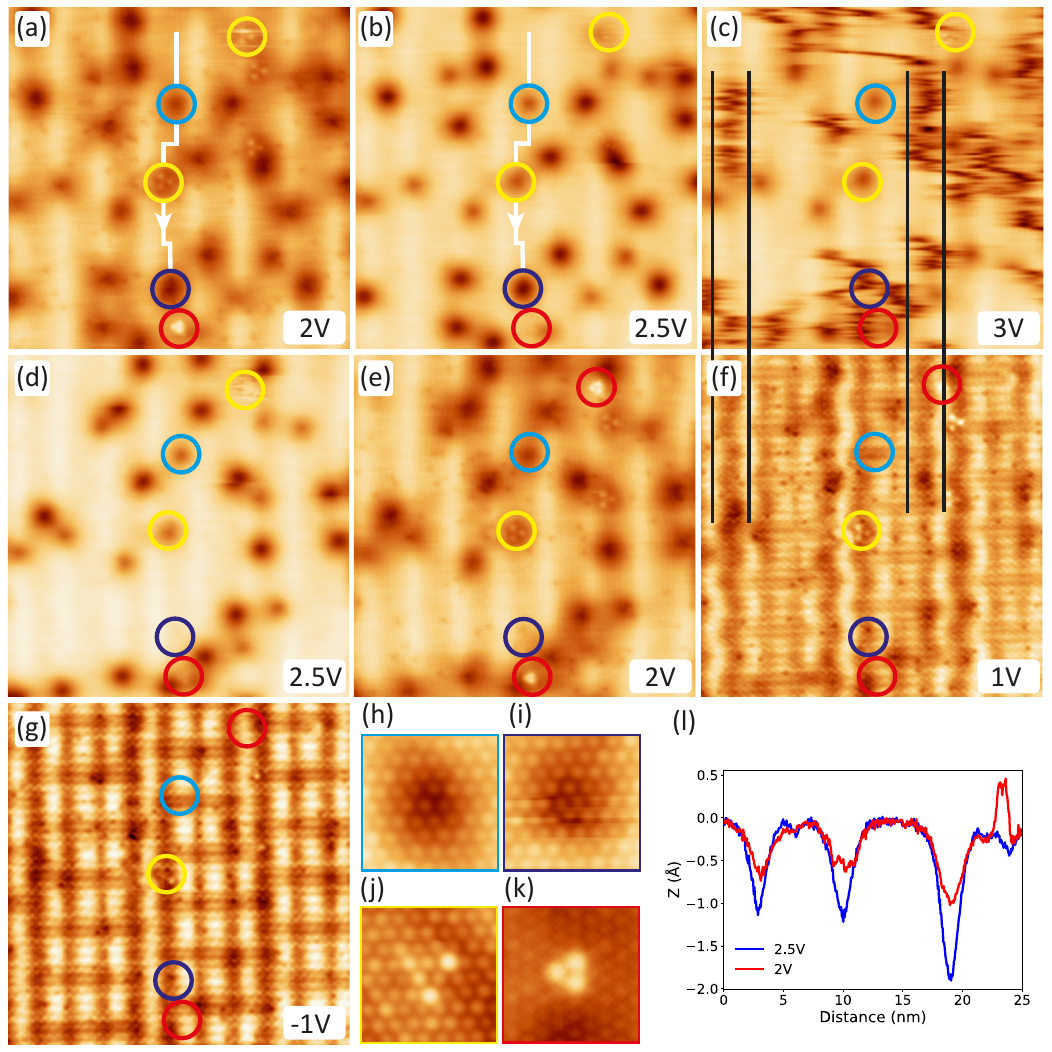}
\caption{Polarons in single layer MnBr$_{2}$ on Gr/Ir(110). (a)-(g) Successive STM topographs of the same sample location acquired at 4.2\,K, with indicated $V_\mathrm{b}$ and $I_\mathrm{t} = 50$\,pA. Polaron types are labeled as follows: light blue circle (type-I immobile), dark blue circle (type-I mobile), yellow circle (type-II 2×2), and red circle (type-II 1×1). Atomically resolved images of (h) type-I immobile, (i) type-I mobile, (j) type-II 2×2, and (k) type-II 1×1 polarons. (l) Height profiles extracted along the white line in (a) and (b). Atomic resolution imaging parameters:(h,k) $V_\mathrm{b} = 2$\,V and $I_\mathrm{t} = 1$\,nA; (i) $V_\mathrm{b} = 1$\,V and $I_\mathrm{t} = 1$\,nA; (j) $V_\mathrm{b} = 2$\,V and $I_\mathrm{t} = 100$\,pA. Image sizes: (a)-(g) 27\,nm $\times$ 27\,nm; (h)-(k) 2.7\,nm $\times$ 2.7\,nm.
}
\label{fig_polaron2}
\end{figure}

To obtain more detailed insight, we imaged an MnBr$_2$ island area with high resolution while varying $V_\mathrm{b}$ between $+3.0$\,V and $-1$\,V, as presented in Figure~\ref{fig_polaron2}. For positive $V_\mathrm{b} \geq 2$\,V [Figure~\ref{fig_polaron2}(a)-(e)] polarons and vertical bright stripes with $\lambda_m = 3.3$\,nm are present. For $V_\mathrm{b} \leq 1$\,V [Figure~\ref{fig_polaron2}(f)-(g)], the contrast of the polarons fades, and the stripes with $\lambda_m = 3.3$\,nm are absent. A new complex pattern evolves, visible already in Figure~\ref{fig_polaron2}(f) and with the strongest contrast in Figure~\ref{fig_polaron2}(g). This pattern is the MnBr$_2$/Gr/Ir(110) super-moiré \cite{safeer25b}. Its dominant Fourier component is the periodicity of the vertical stripes with $\lambda_{sm} = 2.7$\,nm. Although the fine details of the super-moiré are not relevant here, it has more Fourier components, e.g., giving rise to a beating pattern of the stripes with a periodicity of 13.5\,nm and horizontal stripe segments, as outlined in ref. \cite{safeer25b}.  
We conclude that when tunneling is into the conduction band [the CBE is at 1.9\,eV], only the corrugation of the Gr/Ir(110) moiré imprinted on the soft MnBr$_2$ layer modulates the tunneling current. However, with $V_\mathrm{b} < 1.9$\,V, electrons tunnel through the MnBr$_2$ band gap directly to the MnBr$_2$/Gr/Ir(110) interface, where the super-moiré originates \cite{safeer25b}. 

Polarons imaged at $V_\mathrm{b} = +2.0$\,V [Figure~\ref{fig_polaron2}(a)] exhibit distinct shapes and heights. We distinguish four different types of polarons: type-I mobile polarons with the largest apparent depth of $\approx 1$\,\AA~(encircled dark blue), shallower type-I immobile polarons with a depth of $\approx 0.7$\,\AA~(encircled light blue), type-II $2\times2$ polarons (encircled yellow), composed of three protrusions at the corners of a triangle with side length twice the MnBr$_2$ lattice parameter within a shallow depression of about 0.5\,\AA, and type-II $1\times1$ polarons (encircled red) composed of three bright protrusions of height $\approx 0.5$\,\AA~at the corners of a triangle of side length equal to the MnBr$_2$ lattice parameter. Atomically resolved images of the four types of polarons are shown in Figures~\ref{fig_polaron2}(h)-(k). 
The distinction of the four polarons in sets of type-I and type-II is not only based on similarities of their appearance but also in electronic signatures, as we will see below. It is also motivated by similarities to polarons in CoCl$_2$, as will be analyzed in the section \textit{Discussion}. The appearance and visibility of the polarons strongly depend on the sample bias applied, as obvious from the voltage-dependent series of Figure~\ref{fig_polaron2}(a)-(g) and the comparison of the two height profiles for $V_\mathrm{b} = 2$\,V and $V_\mathrm{b} = 2.5$\,V in Figure~\ref{fig_polaron2}(l) along the path indicated in Figures~\ref{fig_polaron2}(a) and (b).

When imaging at $V_\mathrm{b} = +3.0$\,V [Figure~\ref{fig_polaron2}(c)], the fuzzy dark stripes appear, as seen on a large scale in Figure~\ref{Fig_Polaron_presence}(b), indicating substantial mobility during scanning. Only a few polarons remain in position. In fact, type-I mobile polarons are the most abundant species, making up 75\% of all polarons. Upon lowering the sample bias back to $V_\mathrm{b} = 2.5$\,V [Figure~\ref{fig_polaron2}(d)], the mobile polarons become immobile again.

Applying again $V_\mathrm{b}$ = +2.0\,V [Figure~\ref{fig_polaron2}(e)], identical to that in Figure~\ref{fig_polaron2}(a), it is evident that the type-I mobile polaron encircled dark blue in Figure~\ref{fig_polaron2}(a) has changed position, as all other unmarked type-I mobile polarons. All other polarons remained stationary. 
Remarkably, a type-II $2\times2$ polaron encircled yellow in Figure~\ref{fig_polaron2}(a) has transformed into a type-II $1\times1$ polaron encircled red in Figure~\ref{fig_polaron2}(e).

Lowering the sample bias to $V_\mathrm{b} = +1.0$\,V [Figure~\ref{fig_polaron2}(f)] and even further to $V_\mathrm{b} = -1.0$\,V [Figure~\ref{fig_polaron2}(g)] reduces the visibility of individual polarons, while at the same time the super-moiré of MnBr$_2$/Gr/Ir(110) becomes prominent. In Figure~\ref{fig_polaron2}(f), it can be recognized that the remaining type-II $2\times2$ polarons are located at the minima of the super-moiré.

Some of the ridges of the wave pattern with $\lambda_{sm} = 2.7$\,nm of the dominant Fourier component of the super-moiré in Figure~\ref{fig_polaron2}(f) marked by black lines are extended to Figure~\ref{fig_polaron2}(c). Although the fluctuations of the type-I mobile polarons occasionally extend in the horizontal direction over the black lines, it is apparent that the fluctuations are predominantly confined between neighboring lines. The same feature has already been observed in Figure~\ref{Fig_Polaron_presence}(b). 
We therefore conclude that under the driving influence of the STM tip at sufficiently large positive sample bias, type-I mobile polarons experience a repulsive interaction with the maxima of the wave pattern with $\lambda_{sm} = 2.7$\,nm of the MnBr$_2$/Gr/Ir(110) super-moiré.

Type-I mobile polarons also move for large negative $V_\mathrm{b} \leq -3.5$\,V as documented by a more extended data set shown in Figure~S1 in the Supporting Information (SI). We also note that while in Figure~\ref{fig_polaron2} motion of type-I mobile polarons was observed for $V_\mathrm{b} = 3$\,V, motion at lower positive $V_\mathrm{b} = 2$\,V can take place when $I_\mathrm{t}$ is simultaneously increased to reduce the tunneling resistance (see Figure~S2 of SI).

\begin{figure}[ht!]
\includegraphics[width=\textwidth]{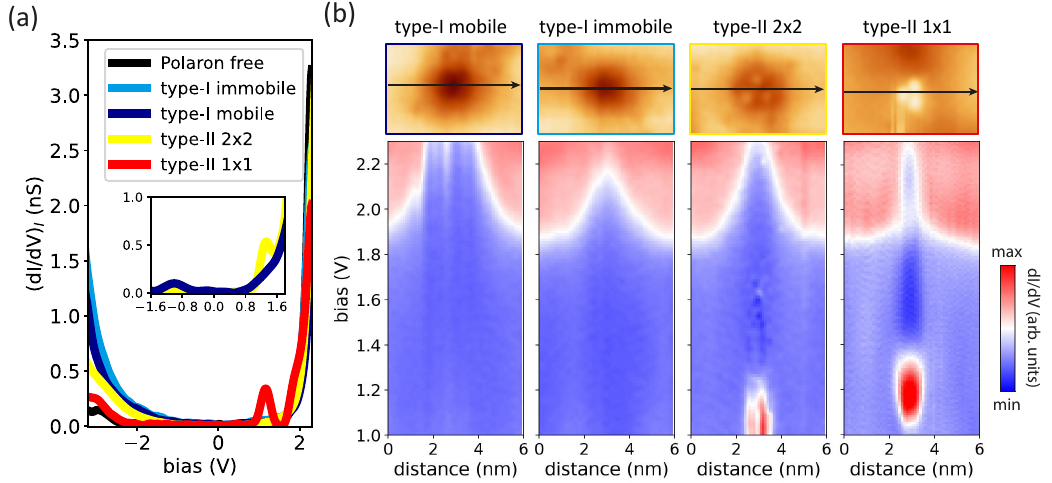}
\caption{Band bending in MnBr$_2$ on Gr/Ir(110) induced by polarons. (a) Constant height $\mathrm{d}I/\mathrm{d}V$ point spectra taken on the polarons indicated, compared to a reference spectrum without a polaron. Inset displays spectra with enhanced sensitivity using a setpoint of smaller tunneling resistance. (b) Lower panel: constant current  $\mathrm{d}I/\mathrm{d}V$ linescans over the type-I mobile, type-I immobile, type-II $2\times2$, and type-II $1\times1$ polaron. Spectra are acquired with $V_{st}$ = +2.3 V, $I_{st}$ = 50\,pA, $f_{mod}$ = 667\,Hz, and $V_{mod}$ = 20\,mV except for inset of (a) where $I_{st}$ = 500\,pA for the spectrum of the type-I mobile polaron and $I_{st}$ = 100\,pA for the spectrum of the type-II $2\times2$ polaron. The path of $\mathrm{d}I/\mathrm{d}V$ linescans through polarons is indicated in the STM topographs in the upper panels of the respective linescan maps. STM imaging parameters: $V_\mathrm{b} = 2$\,V and $I_\mathrm{t} = 100$\,pA and STM images size: 6\,nm $\times$ 4\,nm. }
\label{fig_polaron3} 
\end{figure}

\subsection{Charge state, stability and transformations of polarons}

A polaron is a localized charge dressed by lattice deformation. To substantiate our claim to observe polarons, we took constant height $\mathrm{d}I/\mathrm{d}V$ point spectra on the four different polarons and a reference spectrum on a polaron-free area as shown in Figure~\ref{fig_polaron3}(a).
The spectra recorded on the different polarons exhibit an upward shift in the onset of the valence band, as expected for a localized negative charge. In the conduction band, an upward shift in the onset appears as well, but the situation is less clear. 

Therefore, we applied constant current $\mathrm{d}I/\mathrm{d}V$ spectroscopy at the CBE, which has proven to be extremely sensitive to detecting band edges \cite{Murray19}. Lines of point spectra (briefly line spectra) shown in the lower panel of Figure~\ref{fig_polaron3}(b) were taken along the paths as indicated by the arrows in the STM topographs of the upper panel of Figure~\ref{fig_polaron3}(b). From the line spectra, it is obvious that the polarons cause substantial upward band bending of the CBE, as expected for the presence of a negative charge. The spatial extent of the band bending is of the order of $\pm 2.5$\,nm around the polaron center. The type-II $1\times1$ polaron exhibits the strongest localization of band bending. 

Beyond the band bending, the type-II $2\times2$ and type-II $1\times1$ polarons exhibit in-gap states around $V_\mathrm{b} \sim 1.2$\,V, attributable to localized polaronic states. 
These in-gap states are also visible in the spectra of Figure~\ref{fig_polaron3}(a),  for the type-II $2\times2$ polaron only in the inset, which displays spectra taken with enhanced sensitivity. In the inset spectrum of the type-I mobile polaron, also a polaronic state in the occupied states with $V_\mathrm{b} \sim -1.0$\,V becomes apparent.

\begin{figure}[ht!]
\includegraphics[width=\textwidth]{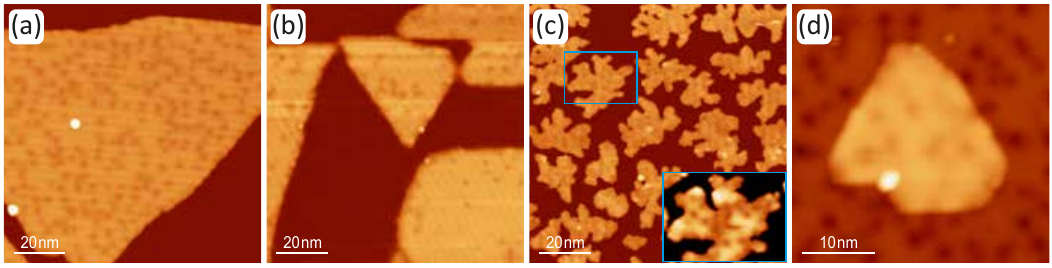}
\caption{Thermal stability and growth independence of polarons in MnBr$_2$ on Gr/Ir(110). (a) Same sample as shown in Figure~\ref{Fig_Polaron_presence}, but STM topograph acquired at 77\,K. (b) STM topograph of a standard MnBr$_2$ sample acquired at 300\,K. (c) Sample after growth at $\approx 80$\,K and taken at 77\,K with unavoidable annealing to $\approx 150$\,K in between due to sample transfer. The inset is a contrast-enhanced zoom of the area marked by a blue rectangle. (d) STM topograph with bilayer MnBr$_2$ island. STM imaging parameters: $V_\mathrm{b} = 2.5$\,V, $I_\mathrm{t} = 20$\,pA. 
Image sizes: (a-c) 100\,nm $\times$ 100\,nm and (d) 35\,nm $\times$ 35\,nm.
}
\label{Fig_Polaron_77K_and_BL} 
\end{figure}

Polarons in MnBr$_2$ on Gr/Ir(110) display a remarkable thermal stability. As shown in Figure~\ref{Fig_Polaron_77K_and_BL}(a) and (b), they are still present at 77\,K and 300\,K, respectively. Their number density decreases with temperature from $(0.060 \pm 0.005)$\,polarons/nm$^{2}$ at 4.2\,K over $(0.040 \pm 0.005)$\,polarons/nm$^{2}$ at 77\,K to $(0.010 \pm 0.005$\,polarons/nm$^{2}$ at 300\,K. The polaron number density at 300\,K has a large relative uncertainty, since at all $V_\mathrm{b}$ where polarons were visible, their interaction with the STM tip was noticeable.

The presence of polarons in single-layer MnBr$_2$ on Gr/Ir(110) is independent of specific growth conditions. Up to now, only polarons grown in samples at 450\,K were presented. Figure~\ref{Fig_Polaron_77K_and_BL}(c) displays islands grown at $\approx 80$\,K, subject to unavoidable annealing to $\approx 150$\,K during UHV sample transfer to the STM, and imaged at 77\,K. Even in the resulting fractal islands, polarons are present, as highlighted by the contrast-enhanced inset of Figure~\ref{Fig_Polaron_77K_and_BL}(c). Their concentration is comparable to that after MnBr$_2$ growth at 450\,K. At variance with this observation is the growth temperature dependence of the polaron concentration in CrBr$_3$ \cite{Cai2025}. However, these variations occur at temperatures equal to or greater than 500\,K. 

Lastly, polarons are also visible in MnBr$_2$ bilayers as shown in the STM topograph of Figure~\ref{Fig_Polaron_77K_and_BL}(d). However, it is not obvious whether these polarons are located in the first or the second MnBr$_2$ layer. We could not investigate polarons in thicker films due to increasing imaging difficulties as thickness increases.

\begin{figure}[ht!]
\includegraphics[width=\textwidth]{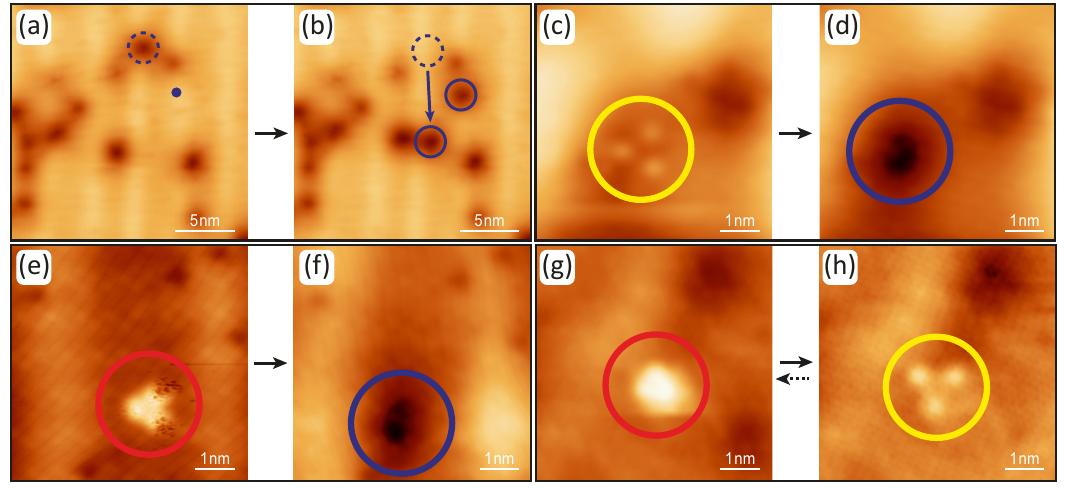}
\caption{Polaron creation and conversion in MnBr$_2$ on Gr/Ir(110). (a) A voltage pulse with $V_\mathrm{b} = 3.5$\,V and $I_\mathrm{t} = 5$\,nA was set at the location indicated by the black dot. (b) The pulse created a type-I mobile polaron under the tip, but also caused a type-I mobile polaron to move (indicated by the arrow, original location encircled in dashed blue). (c) STM topograph of a type-II $2\times2$ polaron. (d) STM topograph of the same area as in (c) after transformation of the type-II $2\times2$ polaron due to scanning into a type-I mobile polaron. (e) STM topographs of type-II $1\times1$ polaron. (f) STM topographs of the same area as in (e) after transformation of a type-II $1\times1$ polaron into a type-I mobile polaron during scanning. (g) STM topograph of type-II $1\times1$ polaron. (h) STM topograph of the same area as in (g) after transformation of the type-II $1\times1$ polaron into a type-II $2\times2$ polaron during scanning.
Scanning parameters: (a,b) $V_\mathrm{b} = 2.5$\,V and $I_\mathrm{t} = 50$\,pA, (c,d) $V_\mathrm{b} = 2$\,V and $I_\mathrm{t} = 50$\,pA, (e,f) $V_\mathrm{b} = 2$\,V and $I_\mathrm{t} = 100$\,pA, and (g,h) $V_\mathrm{b} = 1.5$\,V and $I_\mathrm{t} = 100$\,pA. 
Image sizes: (a,b) 20\,nm $\times$ 20\,nm and (c-h) 6\,nm $\times$ 6\,nm.  
}
\label{Fig_Creation} 
\end{figure}

The creation, annihilation, conversion, and mobility of polarons were systematically investigated for the case of CoCl$_2$ using a wide range of bias voltages with $-3.5\,\mathrm{V} \leq V_\mathrm{b} \leq 4$\,V and up to large tunneling currents $I_\mathrm{t}$ of 80\,nA \cite{cai2023}. Here, a comparable analysis is hindered by the high polaron density and the fact that a voltage pulse simultaneously triggers multiple events in the ensemble. This is visualized in Figure~\ref{Fig_Creation}(a) and (b), where a moderate voltage pulse causes not only the creation of a type-I mobile polaron, but also the motion of a nearby type-I mobile polaron to another position. Generally, type-II polarons easily convert into type-I mobile polarons [Figure~\ref{Fig_Creation}(c-f)], while type-I mobile polarons cannot be converted into any type-II polaron. Type-II polarons also easily intraconvert, as shown for the conversion of a type-II $1\times1$ into a type-II $2\times2$ polaron in Figure~\ref{Fig_Creation}(g) and (h). The opposite process, the conversion of a type-II $2\times2$ polaron into a type-II $1\times1$ polaron, was already observed in the sequence of Figure~\ref{fig_polaron2} and is symbolically indicated by the dashed arrow between Figure~\ref{Fig_Creation}(g) and (h). We tentatively conclude that the type-I mobile polaron is energetically more favorable than the type-II polarons. A type-I immobile polaron could not be created or moved in the parameter space explored by us, that is, for $-3.5\,\mathrm{V} \leq V_\mathrm{b} \leq 3.5$\,V and $I_\mathrm{t} \leq 3$\,nA. This indicates that the type-I immobile polaron is either more stable than the other polarons or better screened against the tip electric field. 

\subsection{Substrate dependence of p}olarons in MnBr$_2$

\begin{figure}[!ht]
\includegraphics[width=\textwidth]{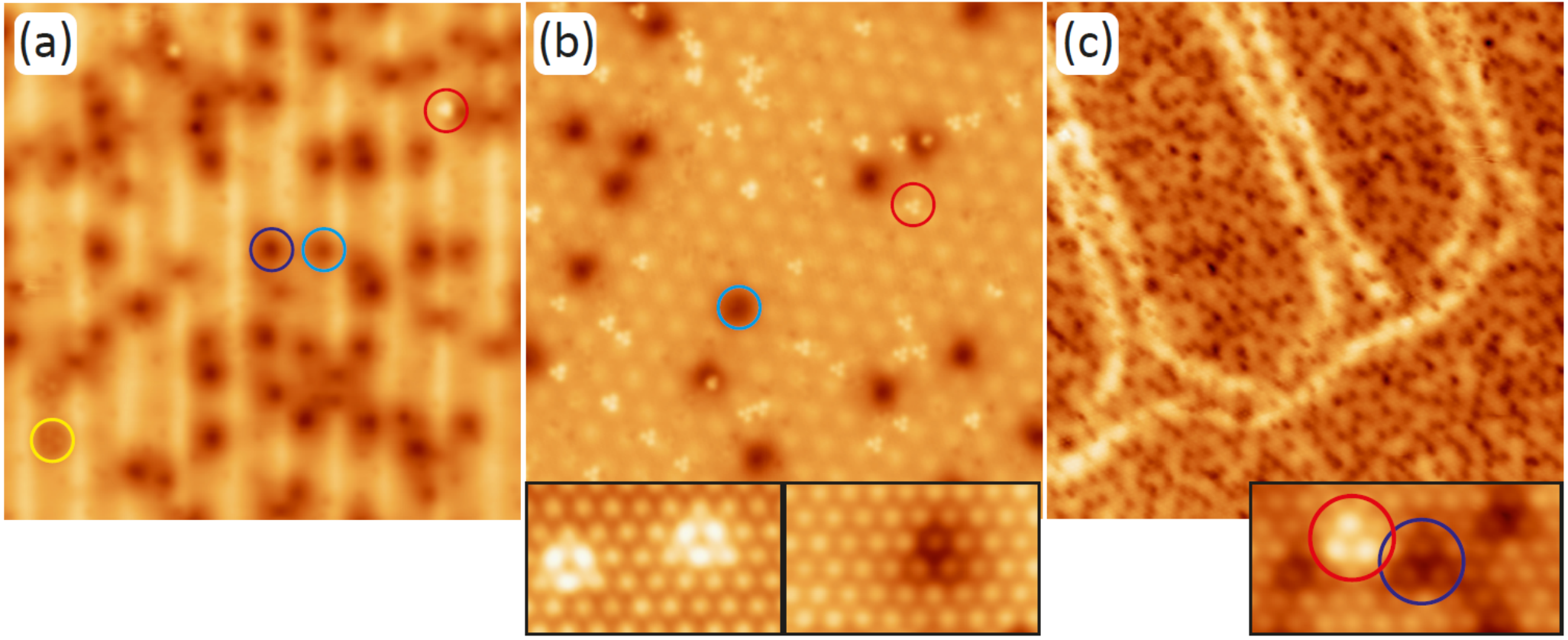}
\caption{Comparison of polarons on different substrates: (a) MnBr$_2$/Gr/Ir(110); (b) MnBr$_2$/Gr/Ir(111); (c) MnBr$_2$/Au(111). Polarons are encircled according to the color code introduced in Figure~\ref{fig_polaron2} and used throughout. Insets in (b) show atomic-resolved type-I immobile and type-II $1\times1$ polarons. In (b), the hexagonal 2.5\,nm Gr/Ir(111) moiré is slightly modulating the height of the MnBr$_2$ layer. Inset (c) shows the type-I mobile and type-II $1\times1$ polaron in high resolution. Bright lines in (c) are due to Au(111) surface reconstruction modulating the height of the MnBr$_2$ layer. STM imaging parameters: (a) $V_\mathrm{b} = 2$\,V, $I_\mathrm{t} = 50$\,pA; (b,c) $V_\mathrm{b} = 2.5$\,V, $I_\mathrm{t} = 20$\,pA. Insets in (b) $V_\mathrm{b} = 1.5$\,V, $I_\mathrm{t} = 100$\,pA, inset in (c) $V_\mathrm{b} = 2.5$\,V, $I_\mathrm{t} = 50$\,pA. Image sizes: (a-c) is 40\,nm $\times$ 40\,nm, insets in (b) 3.4\,nm $\times$ 2\,nm, and inset in (c) 4\,nm $\times$ 2\,nm. 
}
\label{Fig_compare} 
\end{figure}

To clarify the influence of the substrate on the phenomenology of polarons, three single-layer MnBr$_2$ samples were prepared on Gr/Ir(110), Gr/Ir(111), and Au(111). Gr on Ir(110) is structurally intact, but electronically the Dirac cone is absent due to substantial and inhomogeneous chemisorption. The Gr layer displays a sp$^3$-component in binding, making it slightly reactive to adsorption \cite{Kraus22}. Gr on Ir(111) is physisorbed with an almost intact Dirac cone, marginal p-doping, and a clean sp$^2$ character \cite{Busse11}. It can be considered an inert substrate, allowing the preparation of quasi-freestanding materials \cite{Hall18}. Although Au(111) is a noble metal surface, when used for the growth of transition metal chalcogenides, substantial band gap renormalization through hybridization and screening, reconstruction lifting, and the absence of charge density waves due to the interaction of the 2D material with the substrate were observed \cite{Bruix16,Sanders16,Stan19}. 

Using identical growth conditions, Figures\,\ref{Fig_compare}(a)-(c) compare Gr/Ir(110), Gr/Ir(111) and Au(111) as substrates for single-layer MnBr$_2$, respectively. A comparison of larger-scale topographs of the three samples is presented as Figure~S3 in the SI. Already at first glance, Figure~\ref{Fig_compare} makes clear that the differences in polaron phenomenology are substantial. While MnBr$_2$ on Gr/Ir(110) displays in  Figure~\ref{Fig_compare}(a) four different types of polarons, MnBr$_2$ on Gr/Ir(111) in Figure~\ref{Fig_compare}(b) and on Au(111) in  Figure~\ref{Fig_compare}(c) exhibit only two types of polarons. The type-II $1\times1$ polaron is present on all three substrates. On Gr/Ir(111) the type-I immobile polaron is additionally present, whereas the second polaron on Au(111) is tentatively assigned to be the type-I mobile polaron, as discussed in the following. 

Not only the number of polaron species detected, but also the relative and total polaron number densities differ across the three substrates.
The total polaron number density on Au(111) is $(0.6 \pm 0.1)$ \,polarons/nm$^{2}$, about an order of magnitude larger than for Gr/Ir(111) and Gr/Ir(110), both with $(0.060 \pm 0.005)$\,polarons/nm$^{2}$. While on Gr/Ir(110) and Au(111) the type-I mobile polaron is by far the most abundant species, on Gr/Ir(111) it is the type-II $1\times1$ polaron. 

The apparent lateral extension of the type-I polarons in MnBr$_2$ also differs on the three different substrates. It is largest on Gr/Ir(111), slightly smaller on Gr/Ir(110), and considerably smaller on Au(111). One plausible explanation for these different sizes is an increase in screening going from Gr/Ir(111) (low density of states of Gr) to Au(111). 

Although initially absent in MnBr$_2$ on Gr/Ir(111), the type-I mobile polaron can be easily created by conversion of a type-II $1\times1$ polaron during scanning with $V_\mathrm{b} = 2.5$\,V (see Supplementary Note 4 of the SI). In addition, creation of type-I mobile polarons with voltage pulses of $V_\mathrm{b} = 3.5$\,V is straightforward (see Supplementary Note 5 of the SI). Due to the extremely high polaron density in MnBr$_2$ on Au(111), no statement can be made on polaron conversion or creation. Nevertheless, no additional polaron species beyond the two already presented could be identified.

We note that the bright double lines in Figure~\ref{Fig_compare}(c) are dislocation lines of the Au(111) herringbone reconstruction. Their density is much lower than that of pristine Au(111), implying that MnBr$_2$ growth partially lifted the reconstruction. Since chemisorption causes full lifting, the partial presence of the herringbone reconstruction implies only a moderate interaction of the MnBr$_2$ layer with the Au(111) surface, consistent with the integrity and the unchanged lattice parameter of the MnBr$_2$ layer. 

\begin{figure}[!ht]
\includegraphics[width = 0.5\textwidth]{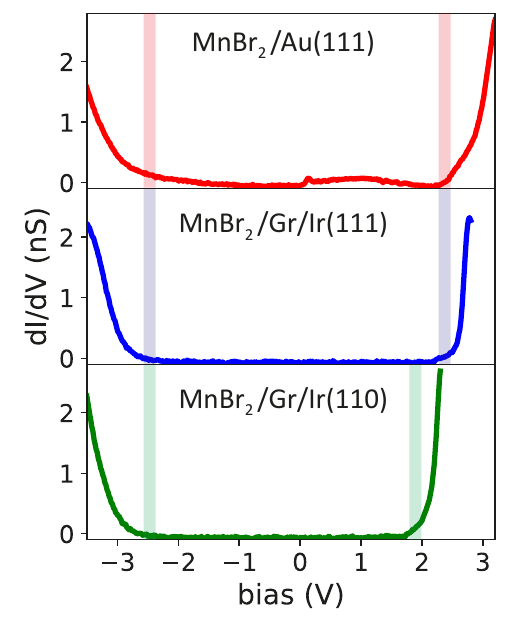}
\caption{Comparison of constant height $\mathrm{d}I/\mathrm{d}V$ point spectra of MnBr$_2$ on different substrates. The conduction and valence band edges are marked by thick colored lines. $\mathrm{d}I/\mathrm{d}V$ spectroscopy parameters: $V_{st}$ = $3.5$\,V for Au(111), $V_{st}$ = $+2.8$\,V for Gr/Ir(111) and $V_{st}$ = $+2.3$\,V for Gr/Ir(110),  $I_{st}$ = 50\,pA, $f_{mod}$ = 667\,Hz, and $V_{mod}$ = 20\,mV.}
\label{Fig_STS_compare} 
\end{figure}

 Figure~\ref{Fig_STS_compare} provides additional insight into the differences in the electronic structure of MnBr$_2$ on the three substrates by a comparison of $\mathrm{d}I/\mathrm{d}V$ point spectra taken in areas without polarons. 
 
 MnBr$_2$ on Gr/Ir(110) and Gr/Ir(111) exhibits well-defined band gaps with the valence band edge located at about $V_\mathrm{b} = -2.6$\,V for both cases. However, for MnBr$_2$ on Gr/Ir(111), the band gap is clearly larger by about 0.4\,eV, with the edge of the conduction band located at $V_\mathrm{b} = 2.2$\,V. For the case of MoS$_2$ a reduction in the size of the band gap (renormalization) due to the interaction with the substrate is well documented \cite{Bruix16}. The smaller band gap size of MnBr$_2$ on Gr/Ir(110) compared to Gr/Ir(111) could therefore be attributed to a band gap renormalization caused by the stronger interaction of MnBr$_2$ with Gr/Ir(110). 
 
For Au(111) as substrate, the measured $\mathrm{d}I/\mathrm{d}V$ signal on MnBr$_2$ displays deviations from a plain band gap. First, a small step-like increase in the $\mathrm{d}I/\mathrm{d}V$ level can be recognized at $V_\mathrm{b} = 0$\,V, which remains until a steep rise at about $V_\mathrm{b} = 2.2$\,V occurs. Second, the valence band edge appears to be less well defined with a gradual rise in the $\mathrm{d}I/\mathrm{d}V$ intensity starting already at -1.5$\,V$ until it takes off at about $V_\mathrm{b} = -2.6$\,V. We attribute these features to the local density of states of the metal surface underneath MnBr$_2$. 
 
 First, a step-like rise in $\mathrm{d}I/\mathrm{d}V$ signal is known for the electron-like parabolic Au(111) surface state. However, the Au(111) surface state has its bottom at $-0.5$\,eV rather than at the Fermi level \cite{Andreev04}. Upward shifts of the surface state by 0.1 - 0.3\,eV due to noble gas adsorption \cite{Andreev04}, growth of Gr \cite{Leicht16}, or growth of single-layer of CoCl$_2$ \cite{Kerschbaumer2025} were reported.
 Since the surface state extends into the vacuum for pristine Au(111), one might speculate that polaron formation in the MnBr$_2$ layer on Au(111) completely emptied the surface state, thereby moving its onset to the Fermi level. The total number density of surface state electrons in pristine Au(111) is at low temperature $0.52$ \,electrons/nm$^{2}$, based on the Fermi wave vectors determined by angle-resolved photoemission spectroscopy \cite{Reinert01,Moue22}. Thus in fact, the polaron number density of $(0.6 \pm 0.1)$\,polaron/nm$^2$ is sufficient to empty the surface state electrons entirely for polaron formation, consistent with observation. Although both number densities agree within the limits of error, it is not evident that the polaron number density is defined by the surface state electrons. Other electrons are also present at the surface of Au(111).

 Second, the density of states of Au is rather flat and dominated by an s-band in the vicinity of the Fermi level \cite{Smith74,Norton78}. The density of states starts to increase at about $1.5$\,eV below the Fermi level, towards lower energies, due to the contribution of the d-band states. We speculate that it is this increase in the Au density of states, which by tunneling through MnBr$_2$ -- transparent in the band gap -- is picked up with the STM.

\begin{figure}[t]
\includegraphics[width=\textwidth]{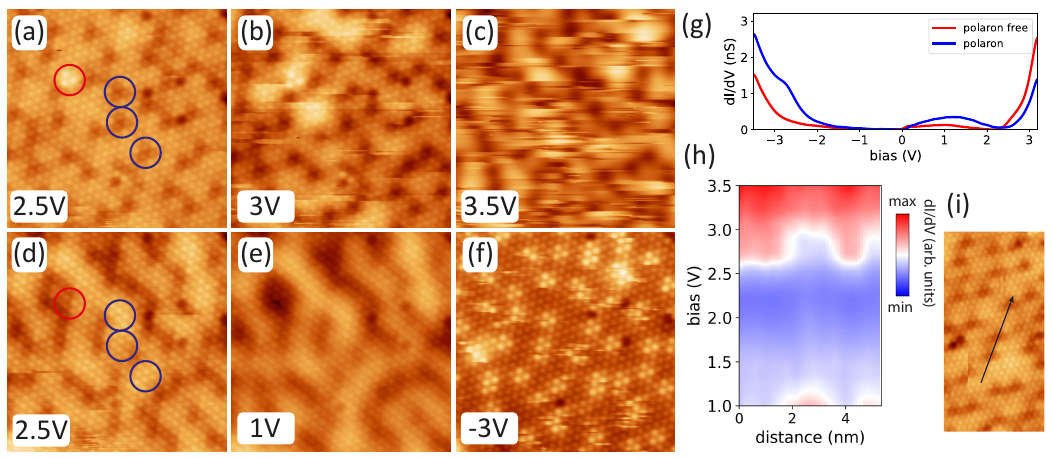}
\caption{Sample bias dependence and electronic structure of polarons in single-layer MnBr$_2$ on Au(111). (a-f) Sequence of high-resolution STM images of single-layer MnBr$_2$ on Au(111) acquired at varying $V_\mathrm{b}$ as indicated and with $I_\mathrm{t} = 50$\,pA. (g) Constant height $\mathrm{d}I/\mathrm{d}V$ point spectrum on type-I mobile polaron compared to the reference spectrum from an area without polaron. (h) Constant current $\mathrm{d}I/\mathrm{d}V$ line scan along the black arrow in the STM image shown in (i). 
STM imaging parameters: (i) $V_\mathrm{b} = +2.5$\,V, $I_\mathrm{t} = 50$\,pA. Constant height and constant current $\mathrm{d}I/\mathrm{d}V$ spectroscopy parameters: $V_{st}$ = $+3.5$\,V, $I_{st}$ = 50\,pA, $f_{mod}$ = 667\,Hz, and $V_{mod}$ = 20\,mV. Image sizes: (a-f) 10\,nm $\times$ 10\,nm, (i) 8\,nm $\times$ 4\,nm. }
\label{fig_Au_polarons} 
\end{figure}
 
The nature of the polarons in MnBr$_2$ on Au(111) is further investigated by bias-dependent STM imaging as shown in  Figure~\ref{fig_Au_polarons}(a-f). In Figure~\ref{fig_Au_polarons}(a) imaged at $V_\mathrm{b} = 2.5$\,V a rare type-II $1\times1$ polaron is encircled red, while three of the type-I mobile polarons are encircled blue. The latter form the overwhelming majority. Increasing the bias to $V_\mathrm{b} = 3$\,V in Figure~\ref{fig_Au_polarons}(b) shows the first indications of polaron mobility. Fuzzy dark stripes appear in Figure~\ref{fig_Au_polarons}(c) imaged with $V_\mathrm{b} = 3.5$\,V. These fuzzy stripes are quite similar to those observed in MnBr$_2$ on Gr/Ir(110) at $V_\mathrm{b} = 3$\,V in Figure~\ref{fig_polaron2}(c). The similar appearance and their mobility at large $V_\mathrm{b}$ justify the assignment of these polarons as type-I mobile polarons. However, a difference from the situation on Gr/Ir(110) becomes obvious at negative $V_\mathrm{b}$: after appearing as somewhat delocalized depressions in Figure~\ref{fig_Au_polarons}(e) at $V_\mathrm{b} = 1$\,V, at $V_\mathrm{b} = -3$\,V as used in Figure~\ref{fig_Au_polarons}(f) the polarons shine-up bright. 
Bright stripes at the polaron locations indicate some mobility also at $V_\mathrm{b} = -3$\,V, again similar to polarons in MnBr$_2$ on Gr/Ir(110) [compare Figure~S1 of the SI]. 

The constant height $\mathrm{d}I/\mathrm{d}V$ spectrum of Figure~\ref{fig_Au_polarons}(g) acquired on a type-I mobile polaron (blue), compared to a reference spectrum from the polaron-free area (red), reveals negative charge localized at the polaron site. This is obvious by the upward shift of both the valence and the CBEs as shown in Figure~\ref{fig_Au_polarons}(g). The polaron spectrum displays near the valence band edge at about $V_\mathrm{b} = -2.8$\,V an additional hump, which makes the polarons shine up bright in Figure~\ref{fig_Au_polarons}(f) and is presumably a polaronic state. Additional intensity between the Fermi level at $V_\mathrm{b} = 0$\,V and the CBE is also observed at the polaron site. This could indicate hybridization between the Au(111) surface state and a polaronic state. A constant current $\mathrm{d}I/\mathrm{d}V$ line scan across type-I mobile polarons in Figure~\ref{fig_Au_polarons}(h) along the path shown in Figure~\ref{fig_Au_polarons}(i) confirms band bending due to negative charge at the polaron sites. It is also obvious from Figure~\ref{fig_Au_polarons}(h) that the band bending is much more localized than for polarons on Gr/Ir(110), returning much faster to the base level of the CBE. We interpret the more localized band bending as well as the more localized appearance of the polarons as a consequence of enhanced screening by the metallic substrate.

For polarons in MnBr$_2$ on Gr/Ir(111), information on voltage-dependent contrast, polaron-induced band bending, polaron creation, and mobility is found in Figures~S4 to S6 of the SI.
 
\section{Discussion}

\textit{Comparison to polarons in single layers of other transition metal halides}

In recent years, polarons in single-layer transition metal halides were observed in several systems: CoCl$_2$ \cite{cai2023,liu2023,Hao24,Song25}, FeCl$_2$ (see Supporting Information of ref.~\cite{cai2023}), NiI$_2$ \cite{miao2025}, and CrBr$_3$ \cite{Cai2025}. For all systems, polarons exhibit a similar global appearance (typically as a nano-sized depression). To focus the discussion, the results from single-layer MnBr$_2$ are compared to single-layer CoCl$_2$, which, up to now, is the most thoroughly investigated among the transition metal halides \cite{cai2023,liu2023,Hao24,Song25,Kerschbaumer2025}. 

Cai et al. \cite{cai2023} observed three different species of polarons in single-layer CoCl$_2$ on highly oriented pyrolytic graphite (HOPG), Liu et al. \cite{liu2023} found two species on HOPG, Hao et al. \cite{Hao24} describe one species (which they consider as dipole solitons) on HOPG, and although Kerschbaumer et al. \cite{Kerschbaumer2025} did not mention polarons they seem to be present in their data of CoCl$_2$ on Au(111) (see below).

Close similarity between the MnBr$_2$ polaron properties and the most detailed observations of Cai et al. \cite{cai2023} is found. They describe three different species of polarons in CoCl$_2$ named type-I (up), type-I (down), and type-II polaron. Due to the apparent similarity of their properties, we tentatively map them onto the type-I mobile, type-I immobile, and type-II $1\times1$ polarons in MnBr$_2$, respectively. In fact, the names of the MnBr$_2$ polarons were chosen deliberately to highlight the similarity to the polarons in CoCl$_2$ and to avoid introducing unnecessary new terminology, without implying anything about the charge position, for which no solid evidence exists in MnBr$_2$.

Type-I polarons in both materials appear as depressions of similar depth and size at $V_\mathrm{b} = 0.6 - 1.0$\,V above the CBE. For both materials, the polarons tend to become invisible in the band gap. For both materials, the polarons with the deeper depression [type I mobile or (up)] can be easily moved with the STM tip. For both materials, type-I polaronic states are hardly visible by STS. Lastly, for both materials, the type-I polarons display upward band bending of the CBE by about 0.5 eV. 

The only apparent difference is that the type-I down polaron can be created, erased, and moved in CoCl$_2$ \cite{cai2023}, while this is not shown here for the type-I immobile polaron in MnBr$_2$. We attribute this difference to the limited range of sample bias $V_\mathrm{b}$ and tunneling current $I_\mathrm{t}$ used in our experiments, rather than to a fundamental difference.

Given the similarity in appearance, bias-dependent changes, mobility, stability, band bending, and visibility of polaronic states, we tentatively conclude that the type-I polarons in both materials are of similar nature.

Cai et al. \cite{cai2023} also observed one species of the type-II polarons, which based on the data provided therein, can be mapped on the type-II $1\times1$ polarons in MnBr$_2$. In both materials, the appearance of the polarons changes as a function of STM imaging voltage from a circular depression to a bright triangular protrusion within a circular shallow depression (compare Figure~\ref{fig_polaron2} and Supplementary Figure~6 of ref.~\cite{cai2023}). Upward bending of the CBE, together with an in-gap polaronic state, is observed.

Referring to the work of Liu et al. \cite{liu2023}, their type-I and type-II polarons in single-layer CoCl$_2$ on HOPG have to be mapped on the type-I (up) and type-I (down) polarons of Cai et al. \cite{cai2023}, and thus correspond to type-I mobile and type-I immobile polarons in MnBr$_2$. 

By STM, Hao et al. and Song et al. \cite{Hao24,Song25} also observed localized depressions in single-layer CoCl$_2$ on HOPG, which they interpret as localized dipoles (solitons). Based on similar phenomenology, including their findings that solitons appear as depressions at positive sample bias and that some of them can be moved by the STM tip, we tentatively conclude that Hao et al. and Song et al. observe the same objects as described by Cai et al. and Liu et al. \cite{cai2023, liu2023}, which correspond to the type-I polarons in MnBr$_2$.

For MnBr$_2$, we rule out the possibility that what we interpret as polarons are instead localized charge dipoles. Figure~\ref{fig_polaron3} documents upward band bending at the polaron location caused by a negative charge. Due to the similar phenomenology in MnBr$_2$ and CoCl$_2$, our finding also supports the polaron interpretation for CoCl$_2$. 

Kerschbaumer et al. \cite{Kerschbaumer2025} investigated the magnetism of single-layer CoCl$_2$ on Au(111) and observed polarons in high number density (compare Figure~1(a) of ref.~\cite{Kerschbaumer2025}), similar to what is found here for single-layer MnBr$_2$ on Au(111). Even band bending and polaronic in-gap states are observed by STS (compare Figure~2 of ref.~\cite{Kerschbaumer2025}). Also, Liu et al. imaged CoCl$_2$ on Au(111), and the STM topographs of Fig. 19 in the Supplementary Information of ref.~\cite{liu2023} display features similar to those observed in MnBr$_2$ on Au(111), which we interpret as polarons. 

Type-II $2\times2$ polarons in MnBr$_2$ cannot be mapped onto any polarons experimentally described in transition metal dihalides so far. From the facile interconversion of type-II $2\times2$ and type-II $1\times1$ polarons [compare Figure~\ref{fig_polaron2}(a) and (e) as well as Figure~\ref{Fig_Creation}(g) and (h)] and their similar appearance as a triangular protrusion, it seems plausible that they are related and distinct from type-I polarons. 
We also note that in DFT calculations conducted in the absence of a substrate, only three different polarons were identified in transition metal halides \cite{Yao2024}, making the description of the fourth polaron species in transition metal halides a relevant finding. Since the type-II $2\times2$ is the only polaron species observed only for one of the three substrates of the MnBr$_2$ layer, its existence and stability appear to be tightly bound to a specific substrate transition metal halide interaction. Therefore, it is understandable that it has not yet been described for other transition metal halide systems and could not be predicted in calculations neglecting the substrate. \\

\textit{Polaron stability} 

Polarons in single-layer MnBr$_2$ form spontaneously without external excitation. They are present in samples grown at 450\,K and even in samples that never experienced a temperature above $\approx 150$\,K [see Figure~\ref{Fig_Polaron_77K_and_BL}(c)]. Their immediate presence in STM topographs under tunneling conditions where no changes in polaron distribution or appearance occur supports this statement. Consequently, their presence is an equilibrium property and reduces the total energy of the system. 


When considering the energetics of polarons, one must distinguish between two quantities. First, the polaron trapping energy $E_\mathrm{tr}$, which is the energy of a localized electron with respect to the conduction band edge energy $E_{\mathrm{CBE}}$. A stable polaron requires $E_\mathrm{tr} < 0$, since otherwise the electron would lower its energy by delocalizing in the conduction band. The second quantity is the charge transition level $E_{\mathrm{ctl}}$ \cite{Elmaslmane18}. $E_{\mathrm{ctl}}$ is a thermodynamic quantity, it denotes the Fermi level position at which the polaron changes its charge state (i.e., discharges). Like $E_\mathrm{tr}$, $E_{\mathrm{ctl}}$ is referred to $E_{\mathrm{CBE}}$ and is therefore negative for electron polarons. For the actual Fermi energy $E_{\mathrm{F}}$ of the system (again referenced to $E_{\mathrm{CBE}}$) it has to hold $E_{\mathrm{ctl}} < E_{\mathrm{F}}$. Otherwise, the trapped polaron would spontaneously discharge. Note that $E_\mathrm{tr}$ is defined for a closed system with an extra electron, while $E_{\mathrm{ctl}}$ is defined for an open system that can exchange electrons with a reservoir (e.g., a metal). Therefore, these energies are usually different. In a case study for TiO$_2$ \cite{Elmaslmane18}, it was found that, in magnitude, $E_{\mathrm{ctl}}$ may exceed $E_{\mathrm{tr}}$ by up to a factor of seven.

One thus expects to measure at least one occupied polaronic state in STS for a polaron species in a material in contact with a metal. Indeed, for type-I mobile polarons, a polaron state is observed at $\approx -1$\,eV on Gr/Ir(110) and at $\approx -2.8$\,eV on Au(111) with respect to the Fermi level.
Assuming somewhat oversimplified that bands are rigid upon a Fermi level shift and subtracting the energy of the conduction band edge $E_\mathrm{CBE}$ results in $E_\mathrm{ctl} \approx -5.0$\,eV for Au(111) and $E_\mathrm{ctl} \approx -2.9$\,eV for Gr/Ir(110).  Although surprisingly large in magnitude, considering that in other systems the magnitude of $E_\mathrm{ctl}$ was smaller by a factor seven compared $E_\mathrm{tr}$ \cite{Elmaslmane18}, one obtains that  $E_\mathrm{tr}$ could reach $-0.4$\,eV to $-0.8$\,eV, which is only somewhat larger in magnitude that the DFT predictions for CoCl$_2$ with $-0.2\,\mathrm{eV} \geq E_\mathrm{tr} \geq -0.7$\,eV \cite{cai2023,liu2023,Yao2024,Liang2025}.

For type-I immobile polarons, no polaronic states could be detected, whereas for type-II polarons only unoccupied polaronic states were observed
(compare Figure~\ref{fig_polaron3}), again similar to the CoCl$_2$ case \cite{cai2023,liu2023, Hao24, Song25}. DFT calculations for CoCl$_2$ polarons generally find more than one polaronic state for a given species \cite{liu2023,cai2023,Yao2024}. Whether any of these states is visible in STS depends on two factors. First, a polaronic state may be inaccessible by STM if its orbital distribution does not extend into the vacuum above the layer, as also pointed out in ref.~\cite{cai2023}. Second, specifically for polaronic states deep in the large bandgap of MnBr$_2$, the tunneling barrier is very high, making it difficult to achieve sufficient spectroscopic sensitivity.\\

\textit{The relation of point defects and polarons}

\begin{figure}[hbt!]
\includegraphics[width=0.5\textwidth]{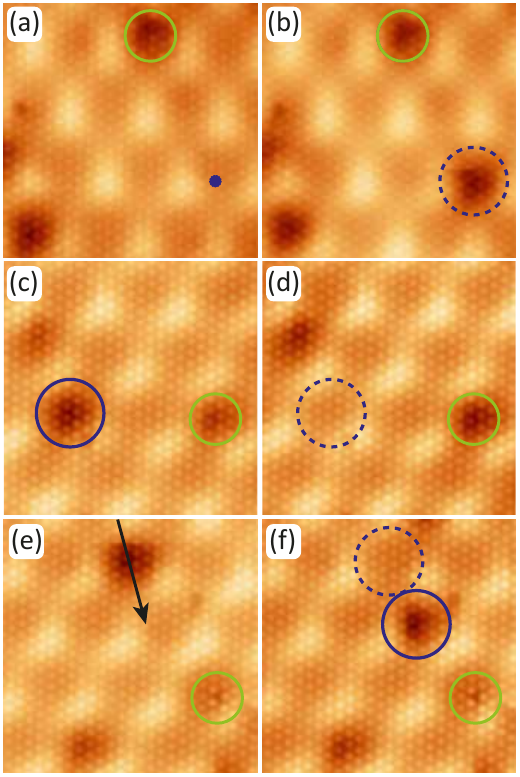}
\caption{Polarons are not structural defects. Atomically resolved STM topographs of type-I mobile polaron creation, annihilation and motion in MnBr$_2$ on Gr/Ir(111). (a), (b): The STM tip was positioned at the defect-free location marked by a blue dot in (a), and a voltage pulse of $V_\mathrm{b} = +5$\,V was applied for 200\,ms. The subsequently acquired STM image in (b) shows a polaron at the pulse location. (c), (d): The STM tip was positioned above the polaron highlighted by the dark blue circle in (c), and a voltage pulse of $V_\mathrm{b} = -3$\,V was applied for 100\,ms. The image (d) acquired afterward shows that the polaron was erased without leaving a visible defect behind. (e), (f): The STM tip was positioned above the polaron highlighted by the dark blue circle in (e). $V_\mathrm{b}$ was increased from $+2$\,V to $+3.5$\,V, and the STM tip was moved over the polaron along the arrow direction. $V_\mathrm{b}$ was then reduced to $+2$\,V and the same location was imaged again as shown in (f). The polaron moved from its original position as defined by the tip motion. Its final position is not influenced by the nearby point defect. The green circles in all STM images highlight a fixed reference location. STM imaging parameters: $V_\mathrm{b} = +2$\,V and $I_\mathrm{t} = 1$\,nA and STM images size: 8.5\,nm $\times$ 8.5\,nm. 
} 
\label{fig_polaron_manipulation} 
\end{figure}
In our STM topographs, we observe point defects which could plausibly be interpreted as Br vacancies. Such point defects are obvious in Figure~\ref{fig_Au_polarons}(f) as dark spots. They are also visible in Figure~\ref{fig_Au_polarons}(a), (b), (d) and (e) always at the same locations. These point defects are immobile, as expected for structural defects. As shown in Figure~S7, despite some variation, these point defects are present in MnBr$_2$ on all three substrates. The point defects are uncharged, as is obvious from the spectroscopic line profile in Figure~S7(d). 

One could thus speculate that the negative charge of Br vacancies detaches from the vacancy location and provides the negative polaron charge. A priori, such a scenario cannot be ruled out. This speculation seems also reasonable, since small polarons in TiO$_2$(110) are pinned to oxygen defects that supply the polaronic charge \cite{Minato09,Setvin14}.   However, the structural defect density in MnBr$_2$ on Au(111) is 0.1\,defects/nm$^{2}$ a factor of six smaller than the corresponding polaron number density.  Moreover, unlike in the TiO$_2$ case, no positional correlation between point defects and polarons can be recognized. We conclude that the absence of positional correlation makes this scenario implausible, and the too small number density for the Au(111) case rules it out as the sole mechanism of charge delivery.

A second, more severe speculation assumes that the polarons described above are in reality charged structural point defects. The following counterarguments rule out this speculation. (i) Due to the strong similarity in the phenomenology of polarons for CoCl$_2$ and MnBr$_2$, the assumption would imply that what is described as polarons in CoCl$_2$ are, in reality, charged structural defects. This seems unreasonable in view of the detailed observations reported \cite{cai2023, liu2023, Yao2024, Hao24, Song25}. 
(ii) As shown in Figures~\ref{fig_polaron_manipulation}(a) and (b), polarons can be created at locations that display a perfect atomic lattice in STM. As is obvious from Figures~\ref{fig_polaron_manipulation}(c) and (d), polarons can be destroyed, leaving behind a perfect atomic lattice. Lastly, as visible in Figures~\ref{fig_polaron_manipulation}(e) and (f), polarons can be moved with the STM tip to locations free of structural defects, leaving no defect behind at the original site. These observations are fatal to the speculation that polarons are just charged structural defects.\\

\textit{Ab initio modeling of polarons in MnBr$_2$}\\
To further investigate small electron polaron formation in MnBr$_2$, we also perform first-principles calculations using density functional theory (DFT). Our approach employs the Heyd–Scuseria–Ernzerhof (HSE) hybrid functional \cite{HSE}\ which predicts the lattice constant (3.88 \AA) and band gap (4.5 eV) of monolayer MnBr$_2$ in good agreement with experiment. Small polaron calculations are performed using a $6 \times 6$ supercell and a 20 \AA\ vacuum gap (see Supplementary Note 8 of the SI for full details). Using the standard HSE functional with 25\% Hartree-Fock (HF) exchange, we were unable to identify a small polaron configuration. However, when the HF exchange is increased to 35\%, we successfully obtain a localized small electron polaron. The symmetry of the localized electronic state is broadly consistent with the experiment (see Figure~S8). The simulated STM image in Figure S8(b) displays three maxima close to the positions of neighboring Br atoms, similar to a type-II $1\times1$ polaron. The associated electronic state is positioned around 0.8\,eV below the conduction band minimum (see Figure~S9), which again fits best to a type-II $1\times1$ polaron [compare Figure \ref{fig_polaron3}(b)]. However, the trapping energy for the polaron with respect to a delocalized electron in the conduction band is small (actually slightly positive at 35\%), and the band gap is increased to 5.4 eV, which is higher than observed in the experiment. Possible origins of this discrepancy could include the influence of the substrate or higher-order many-body effects (both of which are computationally challenging to address). Resolving this issue would require a more extensive dedicated computational study. Nevertheless, the calculations provide some additional support for the formation of small electron polarons in MnBr$_2$.\\

\textit{Consequences of substrate dependence of polarons in MnBr$_2$} 

Although previous ab initio calculations of small polarons in transition metal halides robustly predict polaron formation, polaronic states with their charge distribution, trapping energy, and mobility, the quantitative results depend sensitively on the methodology chosen. For example, Liang et al. \cite{Liang2025} report a variation in the CoCl$_2$ trapping energy by almost a factor of three when results of hybrid functional (HSE) and DFT calculations are compared. 

Substrate effects were only rarely accounted for in previous ab initio modeling of small polaron formation, although it was noted by Yao et al. that interlayer interaction may affect the stability of small CoCl$_2$ polarons \cite{Yao2024}. Despite some charge transfer and a reduced lattice distortion, for small polarons in CoCl$_2$, the orbital character of the HOPG substrate did not cause a change in the major polaron properties in the work of Liu et al. \cite{liu2023}. To the contrary,
DFT calculations for rutile TiO$_2$ interfacing Cu(100) by McKenna  \cite{McKenna16} showed that polaronic charge transfer from Cu to TiO$_2$ leads to the formation of polarons in TiO$_2$ with the polaron trapping energy largest in magnitude next to the Cu/TiO$_2$ interface.

Our experiments, visualized in Figure~\ref{Fig_compare}, make plain that the substrate is decisive for the species and the number density of polarons observed. This finding is not singular. The three investigations of single-layer CoCl$_2$ on HOPG \cite{cai2023,liu2023,Hao24} do not agree on the number of observed polaron (or soliton) species. This may indicate that small differences in substrate preparation or growth procedure are relevant. In CoCl$_2$ on Au(111) \cite{Kerschbaumer2025}, the number density of polarons is much higher than in CoCl$_2$ on HOPG \cite{cai2023,liu2023,Hao24}. This is similar to the situation for MnBr$_2$, where on Au(111) an order of magnitude higher polaron number density is found compared to the two Gr substrates.

These experimental results suggest that ab initio calculations are desirable to investigate and isolate individual substrate-mediated effects such as charge transfer, screening, hybridization, and strain on polaron phenomenology. Given the sensitive dependence of ab initio calculations on the calculation methods and parameters in these correlated systems, together with the supercell limitations for a system involving one or two interfaces (e.g., Ir/Gr/MnBr$_2$), such calculations are extremely demanding and could even miss a relevant feature. Therefore, it may be useful to consider in future experimental work well-designed situations able to separate a single substrate aspect. For instance, through transition metal halide growth on differently doped Gr substrates, the effect of charge transfer and vacuum level alignment could be investigated separately without affecting hybridization, strain, and screening. Such experimental work might provide a hint for the direction of future theoretical investigations of small polaron formation.\\

\textit{Polaron mobility and temperature dependence of number density} 

If present or created by the STM tip, type-I mobile polarons in all three single layer MnBr$_2$ systems become mobile at sample biases above $E_\mathrm{CBE}$, typically by 0.5\,eV to 1\,eV. The same behavior is observed for the corresponding polaron in CoCl$_2$ [type-I (up)]. We interpret the onset of mobility to the population of the conduction band with electrons of sufficient energy and density able to excite phonons. Within this picture, mobility is mainly governed by phonon-assisted polaron hopping \cite{franchini2021, cai2023}, driven by inelastic electron tunneling into the conduction band. 

Within the time scale of 100\,s needed for image acquisition, no polaron mobility is observed at 77\,K [see Figure~\ref{Fig_Polaron_77K_and_BL}(a)]. This timescale corresponds to a hopping frequency $\nu < 10^{-2}$\,s$^{-1}$. 
Transition state theory yields 
\[\nu = \nu_0 e^\frac{-\Delta E}{k_\mathrm{B} T}\]
where $\nu_0$ is a standard prefactor $\nu_0 = 5 \times 10^{12}$ consistent with the optical phonon frequency of $\approx 150$\,cm$^{-1}$ of MnBr$_2$ \cite{Jesus25}, $T$ the temperature, and $k_\mathrm{B}$ the Boltzmann constant. Thus, we can estimate the lower bound for the activation energy $\Delta E$ of phonon-assisted polaron hopping between nearest-neighbor sites to be $\Delta E \geq 0.22$\,eV. For comparison, the highest calculated migration barriers reported for CoCl$_2$ polarons are 117\,meV \cite{cai2023} and 100\,meV \cite{Yao2024}, about a factor of two smaller than the estimate for MnBr$_2$.

The hopping barrier $\Delta E$ must be smaller or equal to the magnitude of $E_\mathrm{tr}$. Equality holds for the case when hopping involves an intermediate delocalization of the electron in the conduction band. Therefore, we conclude $E_\mathrm{tr} \leq - 0.22$\,eV, consistent with our estimate above. Since we identified all four polaron species at 77\,K, this holds for all of them.

Even at 300\,K, immobile polarons are observed, though in a much smaller number density. Given the imaging voltage of $V_\mathrm{b} = 2.5$\,V and their appearance as dark spots, these are type-I polarons. Tip-induced changes such as polaron annihilation and mobility, as well as spectroscopic limitations at 300\,K, make close polaron characterization infeasible. It is plausible that they are type-I immobile polarons. Using the same arguments as above, we conclude that for this polaron species $\Delta E \geq 0.87$\,eV and $E_\mathrm{tr} \leq -0.87$\,eV.

The electric field at the polaron site, defined by tip shape, sample bias, and tip-sample distance, also couples to the trapped electron and the ionic displacement in MnBr$_2$ and may exert a force when the tip is scanned across the polaron \cite{cai2023}. Its lateral component may be relevant for the direction of motion. However, we do not consider the field to be decisive for the onset of polaron mobility for the following reason.
In constant current STM mode, increasing  $V_\mathrm{b}$ from $+2$\,V to $+3$\,V leads to a tip retraction of approximately 3\,\AA~due to the sharp rise in the accessible density of states once the conduction band is reached (see Figure~S10 of the SI). This substantial increase in the tip-sample distance reduces the local electric field and therefore counteracts, likely overcompensates, the direct field enhancement expected from the higher bias, in precisely the bias range where type-I mobile polarons become mobile.

On the observed decrease of the polaron number density with increasing temperature, we can only speculate. If the hopping frequency $\nu$ becomes very high, such that the residence time $\frac{1}{\nu}$ becomes small under the STM tip, polarons are not more detectable in STM topographs. However, besides this visibility issue in STM, it was pointed out that with an increase in the phonon occupation numbers, the self-trapped state of an electron, that is, the small polaron, may abruptly disappear \cite{Emin73}. 

True temperature-dependent STM investigations of polaron concentrations for one and the same sample are desirable to gain quantitative insight into the temperature dependence of polaron mobility and number density. Besides mere imaging, such investigations could be combined with a temperature-dependent analysis of the telegraph noise caused by polaron motion under the STM tip, as, for instance, conducted for a fixed temperature in ref. \cite{cai2023}. 
\\

\textit{Interaction of MnBr$_2$ polarons with the MnBr$_2$/Gr/Ir(110) super-moiré}

A noteworthy observation in our experiments is the interaction of the type-I mobile polaron with the super-moiré of MnBr$_2$/Gr/Ir(110). To our knowledge, an interaction of polarons in a transition-metal dihalide with a moiré has not been previously reported, although this possibility has been checked \cite{liu2023, cai2023}. From Figure~\ref{Fig_Polaron_presence}, Figure~\ref{fig_polaron2}, and Figure~S1(c), we conclude that the vertical stripe maxima of the MnBr$_2$/Gr/Ir(110) super-moiré along the [1$\bar{1}$0]-direction interact repulsively with type-I mobile polarons. The interaction is not so strong that the polarons are confined to a single super-moiré stripe. But the dark fuzzy fluctuations marking the polaron position during scanning make obvious that under the driving influence of the STM tip a type-I mobile polaron typically bounces many times back and forth in a super-moiré valley, before it overcomes a ridge and moves to a neighboring valley. We have to conclude that the super-moiré constitutes a periodic electronic modulation arising from the interaction of the three involved materials, which may also involve structural distortions. While it appears most plausible that a periodic electrostatic potential associated with electronic modulation couples to trapped charge, it cannot be ruled out that periodic distortions of the super-moiré, if present, interact elastically with the deformation cloud of the polaron. 

While a correlation of polarons with the Au(111) herringbone reconstruction is not observed for MnBr$_2$, polarons in single-layer CoCl$_2$ appear to interact with the Au(111) herringbone reconstruction preserved under the layer. Figure~1(a) of ref.~\cite{Kerschbaumer2025} displays a clear correlation between polaron location and the herringbone pattern. 

Together, these observations indicate that substrate-induced potential variations can measurably influence polaron energetics and spatial distribution. Future experiments carefully analyzing the dependence of polaron fluctuations on STM tip voltage and current, as well as testing different electrostatic potential energy or strain landscapes, are in place. They will provide deeper insight into polaron interactions with inhomogeneous electrostatic or elastic environments.

\section{Conclusions}

Single-layers of MnBr$_2$ on Gr/Ir(110) and Gr/Ir(111) host polarons of density $\approx 6 \times 10^{-2}$ \,polarons/nm$^{2}$, while on Au(111) the density is  an order of magnitude larger. Single-layer MnBr$_2$ exhibits four types of polarons on Gr/Ir(110), whereas only two types exist on Gr/Ir(111) and Au(111). All polarons exhibit upward band bending consistent with electron polarons. The lateral extent of the band bending is on the order of 5\,nm on the Gr/Ir substrates, but it is a factor of two smaller on Au(111), presumably due to better screening. 

On the Gr/Ir substrates, type-I mobile polarons can be created and interconverted with both type-II polarons by the STM tip. In contrast, no statements can be made for Au(111) due to the large polaron number density. Type-I mobile polarons move under the influence of a moving STM tip in tunneling contact, when electrons tunnel into states well above the CBE, consistent with phonon-assisted mobility via inelastic tunneling. On Gr/Ir(110), their motion is guided by the superperiodic potential of the super-moiré formed by MnBr$_2$, Gr, and Ir(110).

Of the four different species of polarons identified in MnBr$_2$, three closely resemble those in the paradigmatic CoCl$_2$ on graphite system. This finding suggests that polarons are generic in transition metal halides and are of a similar nature across this material class. At least one polaron type is thermally stable and immobile up to 300\,K, indicating a remarkably high barrier to thermal hopping. Since polarons are invariably present with reproducible number density and distribution of species when imaged gently, the polaron density measured at low temperatures reflects the frozen polaron distribution at some temperature below the growth temperature.  

The key finding of the present work is the strong dependence of polaron phenomenology on substrate. Future experimental work will need to isolate specific substrate effects on polaron formation to uncover the underlying physical principles and guide the future development of theoretical modeling in insulator-on-metal systems.

The interaction of polarons with the super-moiré patterns invites speculation. One could imagine that by patterning substrates, the possible pathways of motion and possible locations of polarons could be laterally predefined at the nanoscale. Together with their thermal stability at 300\,K and the documented ability to create, move, and annihilate polarons, this could open a door for information processing. However, thoughts in this direction are premature without a broader, more precise experimental determination of polaron properties in well-defined systems and their better understanding based on adequate modeling. 

\section{Methods}
The samples were prepared in an ultrahigh vacuum chamber for molecular beam epitaxy with a base pressure below $1 \times 10^{-10}$\,mbar.  Ir(110) is cleaned and prepared in its unreconstructed state using cycles of 1\,keV Ar$^+$ sputtering, flash annealing to 1510\,K, and subsequent cooling to 400\,K in $1\times10^{-7}$\,mbar oxygen pressure. Ir(111) and Au(111) are cleaned through cycles of 1\,keV Ar$^+$ ion sputtering and flash annealing to 1510\,K for Ir(111) or  600\,s annealing to 980\,K for Au(111).

Single crystal Gr was grown by heating clean unreconstructed Ir(110) to 1510\,K and then exposing it to $3\times10^{-7}$\,mbar ethylene for 240\,s \cite{Kraus22}. For Gr growth on Ir(111), the cleaned substrate was first exposed to $1\times10^{-7}$\,mbar ethylene at room temperature for 120\,s and then flash annealed to 1470\,K without ethylene. Subsequently, the sample is exposed for 600\,s to $3\times10^{-7}$\,mbar ethylene at 1370\,K, which completes the growth of a monolayer Gr \cite{Coraux09}. 

MnBr$_{2}$ was grown at 450\,K by sublimation of MnBr$_{2}$ molecules from MnBr$_{2}$ powder in a Knudsen cell heated to 670\,K. The Knudsen cell was placed 8\,cm from the sample, producing a deposition rate of $3 \times 10^{-4}$\,ML/s. Here, a monolayer (ML) corresponds to complete surface coverage by a single layer of MnBr$_{2}$, equivalent to $7.6 \times 10^{18}$\,MnBr$_{2}$ molecules m$^{-2}$ s$^{-1}$. 
  
Once the preparation was completed, the sample was transferred under ultrahigh vacuum into the STM inside a bath cryostat that provides stable STM operation at 77\,K and 4.2\,K. Constant-current topographies were recorded with sample bias $V_\mathrm{b}$ and tunneling current $I_\mathrm{t}$, as detailed in the respective figure captions. The STM images were analyzed and processed (plane subtraction, contrast correction) using WSxM software \cite{Horcas2007}. 
$\mathrm{d}I/\mathrm{d}V$ spectra were obtained at 4.2\,K with stabilization bias $V_{st}$ and stabilization current $I_{st}$ using the standard lock-in technique with modulation frequency $f_{mod}$ and modulation voltage $V_{mod}$, as specified in the captions.

Polaron number densities were determined by analyzing at least 8 different well-resolved topographs for each case. The topographs were taken at bias voltages where polarons are best visible and immobile, typically at sample bias where tunneling is into the conduction band close to the band edge. Except for polarons on Gr/Ir(111) at 300\,K more than 300 polarons were counted. Besides the statistical error due to the limited count, we also estimated systematic uncertainties, which are largest for the Au(111) substrate, where the polarons in MnBr$_2$ are dense.

\section{Acknowledgments}
	\noindent
       Funding from the Deutsche Forschungsgemeinschaft (DFG) through CRC 1238 (project number 277146847, subprojects A01 and B06) is acknowledged. J.F. acknowledges financial support from the DFG through project FI 2624/1-1 (project no. 462692705) within SPP 2137. W.J. acknowledges financial support from the DFG priority program SPP2244 (project no. 535290457). The authors acknowledge useful discussions with Tim Wehling and Achim Rosch.

\section*{Conflict of Interest}
\noindent
The authors declare no conflict of interest.

\section*{Data Availability Statement}
\noindent
The data that support the findings of this study are available from the corresponding author upon reasonable request.


\printbibliography

@article{HSE,
    author = {Krukau, Aliaksandr V. and Vydrov, Oleg A. and Izmaylov, Artur F. and Scuseria, Gustavo E.},
    title = {Influence of the exchange screening parameter on the performance of screened hybrid functionals},
    journal = {The Journal of Chemical Physics},
    volume = {125},
    number = {22},
    pages = {224106},
    year = {2006},
    month = {12},
    issn = {0021-9606},
    doi = {10.1063/1.2404663},
    url = {https://doi.org/10.1063/1.2404663},
}

@article{kresse_efficiency_1996,
	title = {Efficiency of ab-initio total energy calculations for metals and semiconductors using a plane-wave basis set},
	volume = {6},
	issn = {0927-0256},
	url = {https://www.sciencedirect.com/science/article/pii/0927025696000080},
	doi = {10.1016/0927-0256(96)00008-0},
	pages = {15--50},
	number = {1},
	journal = {Computational Materials Science},
	shortjournal = {Computational Materials Science},
	author = {Kresse, G. and Furthmüller, J.},
	urldate = {2025-05-15},
	date = {1996-07-01},
	year = {1996},
}

@article{kresse_efficient_1996,
	title = {Efficient iterative schemes for ab initio total-energy calculations using a plane-wave basis set},
	volume = {54},
	url = {https://link.aps.org/doi/10.1103/PhysRevB.54.11169},
	doi = {10.1103/PhysRevB.54.11169},
	pages = {11169--11186},
	number = {16},
	journal = {Physical Review B},
	shortjournal = {Phys. Rev. B},
	author = {Kresse, G. and Furthmüller, J.},
	urldate = {2025-05-15},
	date = {1996-10-15},
	year = {1996},
}

@article{momma_vesta_2008,
	title = {{VESTA}: a three-dimensional visualization system for electronic and structural analysis},
	volume = {41},
	issn = {0021-8898},
	url = {https://journals.iucr.org/j/issues/2008/03/00/ko5060/},
	doi = {10.1107/S0021889808012016},
	shorttitle = {{VESTA}},
	pages = {653--658},
	number = {3},
	journal = {Journal of Applied Crystallography},
	shortjournal = {J Appl Cryst},
	author = {Momma, K. and Izumi, F.},
	urldate = {2025-05-27},
	date = {2008-06-01},
	year = {2008},
	langid = {english},
}

@article{ganose_sumo_2018,
	title = {sumo: Command-line tools for plotting and analysis of periodic \textit{ab initio} calculations},
	volume = {3},
	issn = {2475-9066},
	url = {https://joss.theoj.org/papers/10.21105/joss.00717},
	doi = {10.21105/joss.00717},
	shorttitle = {sumo},
	pages = {717},
	number = {28},
	journal = {Journal of Open Source Software},
	author = {Ganose, Alex M. and Jackson, Adam J. and Scanlon, David O.},
	urldate = {2025-05-27},
	date = {2018-08-07},
	year = {2018},
	langid = {english},
}

@article{Botana19,
  title = {Electronic structure and magnetism of transition metal dihalides: Bulk to monolayer},
  author = {Botana, A. S. and Norman, M. R.},
  journal = {Phys. Rev. Mater.},
  volume = {3},
  issue = {4},
  pages = {044001},
  numpages = {7},
  year = {2019},
  month = {04},
  publisher = {American Physical Society},
  doi = {10.1103/PhysRevMaterials.3.044001},
}

@article{Horcas2007,
    author = {Horcas, I. and Fernández, R. and Gómez-Rodríguez, J. M. and Colchero, J. and Gómez-Herrero, J. and Baro, A. M.},
    title = {WSXM: A software for scanning probe microscopy and a tool for nanotechnology},
    journal = {Rev. Sci. Instrum.},
    volume = {78},
    number = {1},
    pages = {013705},
    year = {2007},
    month = {01},
    issn = {0034-6748},
    doi = {10.1063/1.2432410},
    url = {https://doi.org/10.1063/1.2432410},

}

@Article{Kulish17,
author = {Kulish, Vadym V. and Huang, Wei},
title  = {Single-layer metal halides MX$_2$ (X = Cl{,} Br{,} I): stability and tunable magnetism from first principles and Monte Carlo simulations},
journal  = {J. Mater. Chem. C},
year  = {2017},
volume  = {5},
issue  = {34},
pages  = {8734},
publisher  = {The Royal Society of Chemistry},
doi  = {10.1039/C7TC02664A},

}

@Article{Li2020,
author = {Li, Xinru and Zhang, Zeying and Zhang, Hongbin},
title  = {High throughput study on magnetic ground states with Hubbard U corrections in transition metal dihalide monolayers},
journal  = {Nanoscale Adv.},
year  = {2020},
volume  = {2},
issue  = {1},
pages  = {495},
publisher  = {RSC},
doi  = {10.1039/C9NA00588A},

}

@article{Luo20,
    author = {Luo, Jia and Xiang, Gang and Tang, Yongliang and Ou, Kai and Chen, Xianmei},
    title = "{The electric and magnetic properties of novel two-dimensional MnBr$_2$ and MnI$_2$ from first-principles calculations}",
    journal = {J. Appl. Phys.},
    volume = {128},
    number = {11},
    pages = {113901},
    year = {2020},
    month = {09},
   issn = {0021-8979},
    doi = {10.1063/5.0015936},
}

@article{Kraus22,
  title = {Single-crystal graphene on Ir(110)},
  author = {Kraus, Stefan and Huttmann, Felix and Fischer, Jeison and Knispel, Timo and Bischof, Ken and Herman, Alexander and Bianchi, Marco and Stan, Raluca-Maria and Holt, Ann Julie and Caciuc, Vasile and Tsukamoto, Shigeru and Wende, Heiko and Hofmann, Philip and Atodiresei, Nicolae and Michely, Thomas},
  journal = {Phys. Rev. B},
  volume = {105},
  pages = {165405},
  numpages = {12},
  year = {2022},
  publisher = {American Physical Society},
  doi = {10.1103/PhysRevB.105.165405},

}

@article{Ronda87,
title = {Photoluminescence and absorption of MnCl$_2$, MnBr$_2$ and MnI$_2$},
journal = {Physica B+C},
volume = {144},
number = {3},
pages = {331-340},
year = {1987},
issn = {0378-4363},
url = {https://www.sciencedirect.com/science/article/pii/0378436387900143},
author = {C.R. Ronda and H.H. Siekman and C. Haas}
}

@article{Sato1994,
title = "Neutron Diffraction Study of Successive Phase Transitions in the Heisenberg Antiferromagnet MnBr2",
author = "Taku Sato and Katsunori Iio and Hiroaki Kadowaki and Hiroji Masuda",
year = "1994",
doi = "10.1143/JPSJ.63.4583",
language = "English",
volume = "63",
pages = "4583--4596",
journal = "J. Phys. Soc. Jpn.",
issn = "0031-9015",
publisher = "The Physical Society of Japan",
number = "12",
}

@article{Wollan58,
  title = {Neutron Diffraction Study of the Magnetic Properties of Mn${\mathrm{Br}}_{2}$},
  author = {Wollan, E. O. and Koehler, W. C. and Wilkinson, M. K.},
  journal = {Phys. Rev.},
  volume = {110},
  issue = {3},
  pages = {638--646},
  numpages = {0},
  year = {1958},
  month = {05},
  publisher = {American Physical Society},
  doi = {10.1103/PhysRev.110.638},
  url = {https://link.aps.org/doi/10.1103/PhysRev.110.638}
}

@article{farge1976,
  title={Optical studies of the magnetic phase diagram of MnBr$_2$},
  author={Farge, Y and R{\'e}gis, M and Royce, BSH},
  journal={J. Phys.},
  volume={37},
  number={5},
  pages={637-644},
  year={1976},
  publisher={Soci{\'e}t{\'e} Fran{\c{c}}aise de Physique},
  DOI= {10.1051/jphys:01976003705063700},
  url= {https://doi.org/10.1051/jphys:01976003705063700}
}

@article{Jesus25,
author = {de Jesus, Guilherme Carlos Carvalho and Vasconcelos, Railson da Concei{\c{c}}ão and do Vale, Lucas Bezerra and de Menezes, Rafael Ferreira and Sprenger, Kayla G. and Gargano, Ricardo},
title = {Electronic and Magnetic Properties of Manganese Bromide Monolayers},
journal = {Langmuir},
volume = {41},
number = {9},
pages = {5877-5883},
year = {2025},
doi = {10.1021/acs.langmuir.4c04499},
URL = {https://doi.org/10.1021/acs.langmuir.4c04499},

}

@article{miao2025,
author = {Mao-Peng Miao  and Nanshu Liu  and Wen-Hao Zhang  and Jian-Wang Zhou  and Dao-Bo Wang  and Cong Wang  and Wei Ji  and Ying-Shuang Fu },
title = {Spin-resolved imaging of atomic-scale helimagnetism in mono- and bilayer NiI$_2$},
journal = {PNAS},
volume = {122},
number = {39},
pages = {e2422868122},
year = {2025},
doi = {10.1073/pnas.2422868122},
URL = {https://www.pnas.org/doi/abs/10.1073/pnas.2422868122},

}

@article{cai2023,
  title={Manipulating single excess electrons in monolayer transition metal dihalide},
  author={Cai, Min and Miao, Mao-Peng and Liang, Yunfan and Jiang, Zeyu and Liu, Zhen-Yu and Zhang, Wen-Hao and Liao, Xin and Zhu, Lan-Fang and West, Damien and Zhang, Shengbai and Fu, Ying-Shuang},
  journal={Nat. Commun.},
  volume={14},
  number={1},
  pages={3691},
  year={2023},
  doi = {10.1038/s41467-023-39360-1},
  URL = {https://doi.org/10.1038/s41467-023-39360-1}, 
  publisher={Nature Publishing Group UK London}
}

@article{liu2023,
  title={Atomic-scale manipulation of single-polaron in a two-dimensional semiconductor},
  author={Liu, Huiru and Wang, Aolei and Zhang, Ping and Ma, Chen and Chen, Caiyun and Liu, Zijia and Zhang, Yi‐Qi and Feng, Baojie and Cheng, Peng and Zhao, Jin and Chen, Lan and Wu, Kehui},
  journal={Nat. Commun.},
  volume={14},
  number={1},
  pages={3690},
  year={2023},
  doi = {10.1038/s41467-023-39361-0},
  URL = {https://doi.org/10.1038/s41467-023-39361-0},
  publisher={Nature Publishing Group UK London}
}

@article{Liang2025,
doi = {10.1088/1361-648X/adb472},
url = {https://dx.doi.org/10.1088/1361-648X/adb472},
year = {2025},
month = {02},
publisher = {IOP Publishing},
volume = {37},
number = {14},
pages = {145601},
author = {Liang, Yunfan and Cai, Min and Peng, Lang and Jiang, Zeyu and West, Damien and Fu, Ying-Shuang and Zhang, Shengbai},
title = {Polaron formation by electron polarization in two-dimensional transition metal halides},
journal = {J. Phys.: Condens. Matter},
}

@article{Yao2024,
  title = {Small polaron dynamics in a two-dimensional magnetic material},
  author = {Yao, Li and Wang, Aolei and Zheng, Qijing and Zhao, Jin},
  journal = {Phys. Rev. B},
  volume = {110},
  issue = {5},
  pages = {054305},
  numpages = {8},
  year = {2024},
  month = {08},
  publisher = {American Physical Society},
  doi = {10.1103/PhysRevB.110.054305},
  url = {https://link.aps.org/doi/10.1103/PhysRevB.110.054305}
}

@article{franchini2021,
  title={Polarons in materials},
  author={Franchini, Cesare and Reticcioli, Michele and Setvin, Martin and Diebold, Ulrike},
  journal={Nat. Rev. Mater.},
  volume={6},
  number={7},
  pages={560--586},
  year={2021},
  publisher={Nature Publishing Group UK London},
  doi = {10.1038/s41578-021-00289-w},
  url = {https://doi.org/10.1038/s41578-021-00289-w}

}

@article{Bo2024,
  title = {Magnetic structure and exchange interactions of transition metal dihalide monolayers: First-principles studies},
  author = {Bo, Xiangyan and Fu, Lei and Wan, Xiangang and Li, Shasha and Pu, Yong},
  journal = {Phys. Rev. B},
  volume = {109},
  issue = {1},
  pages = {014405},
  numpages = {7},
  year = {2024},
  month = {01},
  publisher = {American Physical Society},
  doi = {10.1103/PhysRevB.109.014405},
}

@article{Iio1990,
title = {Successive phase transitions in manganese helimagnets MnX2 (X = I, Br) observed by symmetry breaking birefringence},
journal = {J. Magn. Magn. Mater.},
volume = {90-91},
pages = {265-266},
year = {1990},
issn = {0304-8853},
doi = {https://doi.org/10.1016/S0304-8853(10)80092-7},
author = {K. Iio and H. Masuda and H. Tanaka and K. Nagata},

}

@article{Busse11,
  title = {Graphene on Ir(111): Physisorption with Chemical Modulation},
  author = {Busse, Carsten and Lazi\'{c}, Predrag and Djemour, Rabie and Coraux, Johann and Gerber, Timm and Atodiresei, Nicolae and Caciuc, Vasile and Brako, Radovan and N'Diaye, Alpha T. and Bl\"ugel, Stefan and Zegenhagen, J\"org and Michely, Thomas},
  journal = {Phys. Rev. Lett.},
  volume = {107},
  issue = {3},
  pages = {036101},
  numpages = {4},
  year = {2011},
  month = {07},
  publisher = {American Physical Society},
  doi = {10.1103/PhysRevLett.107.036101},
  url = {https://link.aps.org/doi/10.1103/PhysRevLett.107.036101}
}

@article{Hall18,
doi = {10.1088/2053-1583/aaa1c5},
url = {https://doi.org/10.1088/2053-1583/aaa1c5},
year = {2018},
month = {01},
publisher = {IOP Publishing},
volume = {5},
number = {2},
pages = {025005},
author = {Hall, Joshua and Pielić, Borna and Murray, Clifford and Jolie, Wouter and Wekking, Tobias and Busse, Carsten and Kralj, Marko and Michely, Thomas},
title = {Molecular beam epitaxy of quasi-freestanding transition metal disulphide monolayers on van der Waals substrates: a growth study},
journal = {2D Mater.}
}

@article{Coraux09,
doi = {10.1088/1367-2630/11/2/023006},
url = {https://dx.doi.org/10.1088/1367-2630/11/2/023006},
year = {2009},
month = {2},
publisher = {},
volume = {11},
number = {2},
pages = {023006},
author = {Coraux, Johann and T N'Diaye, Alpha and Engler, Martin and Busse, Carsten and Wall, Dirk and Buckanie, Niemma and Meyer zu Heringdorf, Frank-J and van Gastel, Raoul and Poelsema, Bene and Michely, Thomas},
title = {Growth of graphene on Ir(111)},
journal = {New J. Phys.}
}

@article{Hao24,
author = {Hao, Feng and Song, Rui and Yang, Jie and Shen, Jian and Ernst, Arthur and Yin, Lifeng and Wang, Zhongjie and Gao, Chunlei},
title = {Manipulative Single Electric Dipole with Spontaneous Translational Symmetry Breaking in a Two-Dimensional Crystal},
journal = {Nano Lett.},
volume = {24},
number = {44},
pages = {14042-14049},
year = {2024},
doi = {10.1021/acs.nanolett.4c03858},
URL = {https://doi.org/10.1021/acs.nanolett.4c03858}
}

@article{Murray19,
  title = {Comprehensive tunneling spectroscopy of quasifreestanding ${\mathrm{MoS}}_{2}$ on graphene on Ir(111)},
  author = {Murray, Clifford and Jolie, Wouter and Fischer, Jeison A. and Hall, Joshua and van Efferen, Camiel and Ehlen, Niels and Gr\"uneis, Alexander and Busse, Carsten and Michely, Thomas},
  journal = {Phys. Rev. B},
  volume = {99},
  issue = {11},
  pages = {115434},
  numpages = {8},
  year = {2019},
  month = {3},
  publisher = {American Physical Society},
  doi = {10.1103/PhysRevB.99.115434},
  url = {https://link.aps.org/doi/10.1103/PhysRevB.99.115434}
}

@article{Song25,
doi = {10.1088/1674-1056/adbee8},
url = {https://dx.doi.org/10.1088/1674-1056/adbee8},
year = {2025},
month = {5},
publisher = {Chinese Physical Society and IOP Publishing Ltd},
volume = {34},
number = {5},
pages = {056802},
author = {Song, Rui and Hao, Feng and Yang, Jie and Yin, Lifeng and Shen, Jian},
title = {Surface solitonic charge distribution on 2D materials investigated using Kelvin probe force microscopy technique based on qplus atomic force microscopy},
journal = {Chin. Phys. B}
}

@Article{Cai2025,
author={Cai, Min
and Jiang, Zeyu
and Liao, Wen-Ao
and Qin, Hao-Jun
and Zhang, Wen-Hao
and Zhou, Jian-Wang
and Liu, Li-Si
and Liang, Yunfan
and West, Damien
and Zhang, Shengbai
and Fu, Ying-Shuang},
title={Apparent charge reduction in multipolarons crafted one-by-one in monolayer CrBr$_3$},
journal={Nat. Commun.},
year={2025},
day={02},
volume={16},
number={1},
pages={7117},
issn={2041-1723},
doi={10.1038/s41467-025-62552-w},
url={https://doi.org/10.1038/s41467-025-62552-w}
}

@article{Kerschbaumer2025,
author = {Kerschbaumer, Samuel and Hadjadj, Sebastien Elie and Aguirre-Ba{\~n}os, Andrea and Longo, Danilo and Pinar Sol{\'e}, Andr{\'e}s and Stetsovych, Oleksandr and Candia, Adriana Elizabet and Angulo-Portugal, Paula and Caldevilla, David and Choueikani, Fadi and Corso, Martina and Serrate, David and Lobo-Checa, Jorge and Jelínek, Pavel and Ilyn, Maxim and Rogero, Celia},
title = {Strong In-plane Magnetic Anisotropy in Semiconducting Monolayer CoCl$_2$},
journal = {ACS Nano},
volume = {19},
number = {22},
pages = {20693-20701},
year = {2025},
doi = {10.1021/acsnano.5c02175},
URL = {https://doi.org/10.1021/acsnano.5c02175}
}

@article{Bruix16,
  title = {Single-layer ${\text{MoS}}_{2}$ on Au(111): Band gap renormalization and substrate interaction},
  author = {Bruix, Albert and Miwa, Jill A. and Hauptmann, Nadine and Wegner, Daniel and Ulstrup, S\o{}ren and Gr\o{}nborg, Signe S. and Sanders, Charlotte E. and Dendzik, Maciej and Grubi\ifmmode \check{s}\else \v{s}\fi{}i\ifmmode \acute{c}\else \'{c}\fi{} \ifmmode \check{C}\else \v{C}\fi{}abo, Antonija and Bianchi, Marco and Lauritsen, Jeppe V. and Khajetoorians, Alexander A. and Hammer, Bj\o{}rk and Hofmann, Philip},
  journal = {Phys. Rev. B},
  volume = {93},
  issue = {16},
  pages = {165422},
  numpages = {10},
  year = {2016},
  month = {04},
  publisher = {American Physical Society},
  doi = {10.1103/PhysRevB.93.165422},
  url = {https://link.aps.org/doi/10.1103/PhysRevB.93.165422}
}

@article{Stan19,
  title = {Epitaxial single-layer ${\mathrm{NbS}}_{2}$ on Au(111): Synthesis, structure, and electronic properties},
  author = {Stan, Raluca-Maria and Mahatha, Sanjoy K. and Bianchi, Marco and Sanders, Charlotte E. and Curcio, Davide and Hofmann, Philip and Miwa, Jill A.},
  journal = {Phys. Rev. Mater.},
  volume = {3},
  issue = {4},
  pages = {044003},
  numpages = {6},
  year = {2019},
  month = {04},
  publisher = {American Physical Society},
  doi = {10.1103/PhysRevMaterials.3.044003},
  url = {https://link.aps.org/doi/10.1103/PhysRevMaterials.3.044003}
}

@article{Sanders16,
  title = {Crystalline and electronic structure of single-layer ${\mathrm{TaS}}_{2}$},
  author = {Sanders, Charlotte E. and Dendzik, Maciej and Ngankeu, Arlette S. and Eich, Andreas and Bruix, Albert and Bianchi, Marco and Miwa, Jill A. and Hammer, Bj\o{}rk and Khajetoorians, Alexander A. and Hofmann, Philip},
  journal = {Phys. Rev. B},
  volume = {94},
  issue = {8},
  pages = {081404},
  numpages = {5},
  year = {2016},
  month = {08},
  publisher = {American Physical Society},
  doi = {10.1103/PhysRevB.94.081404},
  url = {https://link.aps.org/doi/10.1103/PhysRevB.94.081404}
}

@article{Andreev04,
  title = {Adsorbed rare-gas layers on Au(111): Shift of the Shockley surface state studied with ultraviolet photoelectron spectroscopy and scanning tunneling spectroscopy},
  author = {Andreev, Thomas and Barke, Ingo and H\"ovel, Heinz},
  journal = {Phys. Rev. B},
  volume = {70},
  issue = {20},
  pages = {205426},
  numpages = {13},
  year = {2004},
  month = {11},
  publisher = {American Physical Society},
  doi = {10.1103/PhysRevB.70.205426},
  url = {https://link.aps.org/doi/10.1103/PhysRevB.70.205426}
}

@article{Norton78,
title = {High resolution photoemission study of the physisorption and chemisorption of CO on copper and gold},
journal = {Surf. Sci.},
volume = {72},
number = {1},
pages = {33-44},
year = {1978},
issn = {0039-6028},
doi = {https://doi.org/10.1016/0039-6028(78)90375-8},
url = {https://www.sciencedirect.com/science/article/pii/0039602878903758},
author = {P.R Norton and R.L Tapping and J.W Goodale}
}

@article{McKenna16,
  title = {Effect of polaronic charge transfer on band alignment at the $\mathrm{Cu}/{\mathrm{TiO}}_{2}$ interface},
  author = {McKenna, Keith P.},
  journal = {Phys. Rev. B},
  volume = {94},
  issue = {15},
  pages = {155147},
  numpages = {5},
  year = {2016},
  month = {10},
  publisher = {American Physical Society},
  doi = {10.1103/PhysRevB.94.155147},
  url = {https://link.aps.org/doi/10.1103/PhysRevB.94.155147}
}

@Article{Sio23,
author={Sio, Weng Hong
and Giustino, Feliciano},
title={Polarons in two-dimensional atomic crystals},
journal={Nat. Phys.},
year={2023},
month={05},
day={01},
volume={19},
number={5},
pages={629-636},
doi={10.1038/s41567-023-01953-4},
url={https://doi.org/10.1038/s41567-023-01953-4}
}

@article{Minato09,
    author = {Minato, Taketoshi and Sainoo, Yasuyuki and Kim, Yousoo and Kato, Hiroyuki S. and Aika, Ken-ichi and Kawai, Maki and Zhao, Jin and Petek, Hrvoje and Huang, Tian and He, Wei and Wang, Bing and Wang, Zhuo and Zhao, Yan and Yang, Jinlong and Hou, J. G.},
    title = {The electronic structure of oxygen atom vacancy and hydroxyl impurity defects on titanium dioxide (110) surface},
    journal = {J. Chem. Phys.},
    volume = {130},
    number = {12},
    pages = {124502},
    year = {2009},
    month = {03},
    issn = {0021-9606},
    doi = {10.1063/1.3082408},
    url = {https://doi.org/10.1063/1.3082408}
   }

@article{Setvin14,
  title = {Direct View at Excess Electrons in ${\mathrm{TiO}}_{2}$ Rutile and Anatase},
  author = {Setvin, Martin and Franchini, Cesare and Hao, Xianfeng and Schmid, Michael and Janotti, Anderson and Kaltak, Merzuk and Van de Walle, Chris G. and Kresse, Georg and Diebold, Ulrike},
  journal = {Phys. Rev. Lett.},
  volume = {113},
  issue = {8},
  pages = {086402},
  numpages = {5},
  year = {2014},
  month = {08},
  publisher = {American Physical Society},
  doi = {10.1103/PhysRevLett.113.086402},
  url = {https://link.aps.org/doi/10.1103/PhysRevLett.113.086402}
}

@article{Skreekumar25,
    author = {Sreekumar, Sreehari and Kocán, Pavel and Setvin, Martin},
    title = {Tracking polarons in real space by STM/AFM},
    journal = {Appl. Phys. Lett.},
    volume = {127},
    number = {14},
    pages = {140502},
    year = {2025},
    month = {10},
   issn = {0003-6951},
    doi = {10.1063/5.0288242},
    url = {https://doi.org/10.1063/5.0288242}
}

@article{Yim16,
  title = {Engineering Polarons at a Metal Oxide Surface},
  author = {Yim, C. M. and Watkins, M. B. and Wolf, M. J. and Pang, C. L. and Hermansson, K. and Thornton, G.},
  journal = {Phys. Rev. Lett.},
  volume = {117},
  issue = {11},
  pages = {116402},
  numpages = {5},
  year = {2016},
  month = {09},
  publisher = {American Physical Society},
  doi = {10.1103/PhysRevLett.117.116402},
  url = {https://link.aps.org/doi/10.1103/PhysRevLett.117.116402}
}

@article{Reticcioli17,
  title = {Polaron-Driven Surface Reconstructions},
  author = {Reticcioli, Michele and Setvin, Martin and Hao, Xianfeng and Flauger, Peter and Kresse, Georg and Schmid, Michael and Diebold, Ulrike and Franchini, Cesare},
  journal = {Phys. Rev. X},
  volume = {7},
  issue = {3},
  pages = {031053},
  numpages = {10},
  year = {2017},
  month = {09},
  publisher = {American Physical Society},
  doi = {10.1103/PhysRevX.7.031053},
  url = {https://link.aps.org/doi/10.1103/PhysRevX.7.031053}
}

@article{Redondo24,
author = {Jesus Redondo  and Michele Reticcioli  and Vit Gabriel  and Dominik Wrana  and Florian Ellinger  and Michele Riva  and Giada Franceschi  and Erik Rheinfrank  and Igor Sokolović  and Zdenek Jakub  and Florian Kraushofer  and Aji Alexander  and Eduard Belas  and Laerte L. Patera  and Jascha Repp  and Michael Schmid  and Ulrike Diebold  and Gareth S. Parkinson  and Cesare Franchini  and Pavel Kocan  and Martin Setvin },
title = {Real-space investigation of polarons in hematite Fe$_2$O$_3$},
journal = {Sci. Adv.},
volume = {10},
number = {44},
pages = {eadp7833},
year = {2024},
doi = {10.1126/sciadv.adp7833},
URL = {https://www.science.org/doi/abs/10.1126/sciadv.adp7833},

}

@article{Yim24,
author = {Yim, Chi-Ming and Allan, Michael and Pang, Chi Lun and Thornton, Geoff},
title = {Scanning Tunneling Microscopy Visualization of Polaron Charge Trapping by Hydroxyls on TiO$_2$(110)},
journal = {J. Phys. Chem. C},
volume = {128},
number = {33},
pages = {14100-14106},
year = {2024},
doi = {10.1021/acs.jpcc.4c03751},
URL = {https://doi.org/10.1021/acs.jpcc.4c03751},

}

@Article{Birschitzky24,
author={Birschitzky, Viktor C.
and Sokolovi{\'{c}}, Igor
and Prezzi, Michael
and Palot{\'a}s, Kriszti{\'a}n
and Setv{\'i}n, Martin
and Diebold, Ulrike
and Reticcioli, Michele
and Franchini, Cesare},
title={Machine learning-based prediction of polaron-vacancy patterns on the TiO$_2$(110) surface},
journal={npj Comput. Mater.},
year={2024},
month={05},
day={06},
volume={10},
number={1},
pages={89},
issn={2057-3960},
doi={10.1038/s41524-024-01289-4},
url={https://doi.org/10.1038/s41524-024-01289-4}
}

@article{Reticcioli19,
  title = {Interplay between Adsorbates and Polarons: CO on Rutile ${\mathrm{TiO}}_{2}(110)$},
  author = {Reticcioli, Michele and Sokolovi\ifmmode \acute{c}\else \'{c}\fi{}, Igor and Schmid, Michael and Diebold, Ulrike and Setvin, Martin and Franchini, Cesare},
  journal = {Phys. Rev. Lett.},
  volume = {122},
  issue = {1},
  pages = {016805},
  numpages = {6},
  year = {2019},
  month = {01},
  publisher = {American Physical Society},
  doi = {10.1103/PhysRevLett.122.016805},
  url = {https://link.aps.org/doi/10.1103/PhysRevLett.122.016805}
}

@article{Chen18,
author = {Chen, Chaoyu and Avila, Jos{\'e} and Wang, Shuopei and Wang, Yao and Mucha-Kruczyński, Marcin and Shen, Cheng and Yang, Rong and Nosarzewski, Benjamin and Devereaux, Thomas P. and Zhang, Guangyu and Asensio, Maria Carmen},
title = {Emergence of Interfacial Polarons from Electron–Phonon Coupling in Graphene/h-BN van der Waals Heterostructures},
journal = {Nano Lett.},
volume = {18},
number = {2},
pages = {1082-1087},
year = {2018},
doi = {10.1021/acs.nanolett.7b04604},
URL = {https://doi.org/10.1021/acs.nanolett.7b04604},

}

@Article{Garcia19,
author={Garcia-Goiricelaya, Peio
and Lafuente-Bartolome, Jon
and Gurtubay, Idoia G.
and Eiguren, Asier},
title={Long-living carriers in a strong electron--phonon interacting two-dimensional doped semiconductor},
journal={Commun. Phys.},
year={2019},
day={15},
volume={2},
number={1},
pages={81},
issn={2399-3650},
doi={10.1038/s42005-019-0182-0},
url={https://doi.org/10.1038/s42005-019-0182-0}
}

@article{VanEfferen25,
  title = {Inelastic Tunneling into Multipolaronic Bound States in Single-Layer ${\mathrm{MoS}}_{2}$},
  author = {van Efferen, Camiel and P\"atzold, Laura and Tounsi, Tfyeche Y. and Schobert, Arne and Winter, Michael and 't Veld, Yann in and Georger, Mark and Safeer, Affan and Kr\"amer, Christian and Fischer, Jeison and Berges, Jan and Michely, Thomas and Mozara, Roberto and Wehling, Tim and Jolie, Wouter},
  journal = {Phys. Rev. X},
  volume = {15},
  issue = {3},
  pages = {031030},
  numpages = {16},
  year = {2025},
  month = {7},
  publisher = {American Physical Society},
  doi = {10.1103/l8lg-ny6m},
  url = {https://link.aps.org/doi/10.1103/l8lg-ny6m}
}

@article{Miyata17,
author = {Kiyoshi Miyata  and Daniele Meggiolaro  and M. Tuan Trinh  and Prakriti P. Joshi  and Edoardo Mosconi  and Skyler C. Jones  and Filippo De Angelis  and X.-Y. Zhu },
title = {Large polarons in lead halide perovskites},
journal = {Sci. Adv.},
volume = {3},
number = {8},
pages = {e1701217},
year = {2017},
doi = {10.1126/sciadv.1701217},
URL = {https://www.science.org/doi/abs/10.1126/sciadv.1701217},

}

@article{Ghosh20,
author = {Ghosh, Dibyajyoti and Welch, Eric and Neukirch, Amanda J. and Zakhidov, Alex and Tretiak, Sergei},
title = {Polarons in Halide Perovskites: A Perspective},
journal = {J. Phys. Chem. Lett.},
volume = {11},
number = {9},
pages = {3271-3286},
year = {2020},
doi = {10.1021/acs.jpclett.0c00018},
URL = {https://doi.org/10.1021/acs.jpclett.0c00018},

}

@Inbook{Reticcioli19b,
author="Reticcioli, Michele
and Diebold, Ulrike
and Kresse, Georg
and Franchini, Cesare",
editor="Andreoni, Wanda
and Yip, Sidney",
title="Small Polarons in Transition Metal Oxides",
bookTitle="Handbook of Materials Modeling: Applications: Current and Emerging Materials",
year="2019",
publisher="Springer International Publishing",
address="Cham",
pages="1--39",
isbn="978-3-319-50257-1",
doi="10.1007/978-3-319-50257-1_52-1",
url="https://doi.org/10.1007/978-3-319-50257-1_52-1"
}

@article{Rettie16,
author = {Rettie, Alexander J. E. and Chemelewski, William D. and Emin, David and Mullins, C. Buddie},
title = {Unravelling Small-Polaron Transport in Metal Oxide Photoelectrodes},
journal = {J. Phys. Chem. Lett},
volume = {7},
number = {3},
pages = {471-479},
year = {2016},
doi = {10.1021/acs.jpclett.5b02143},
note ={PMID: 26758715},
URL = {https://doi.org/10.1021/acs.jpclett.5b02143},
}

@article{Cheng23,
author = {Cheng, Cheng and Zhou, Zhaohui and Long, Run},
title = {Time-Domain View of Polaron Dynamics in Metal Oxide Photocatalysts},
journal = {J. Phys. Chem. Lett.},
volume = {14},
number = {49},
pages = {10988-10998},
year = {2023},
doi = {10.1021/acs.jpclett.3c02869},
URL = {https://doi.org/10.1021/acs.jpclett.3c02869},

}

@article{Tanner22,
author = {Tanner, Alex J. and Thornton, Geoff},
title = {TiO$_2$ Polarons in the Time Domain: Implications for Photocatalysis},
journal = {J. Phys. Chem. Lett.},
volume = {13},
number = {2},
pages = {559-566},
year = {2022},
doi = {10.1021/acs.jpclett.1c03677},
URL = {https://doi.org/10.1021/acs.jpclett.1c03677},

}

@Article{Kang18,
author={Kang, Mingu
and Jung, Sung Won
and Shin, Woo Jong
and Sohn, Yeongsup
and Ryu, Sae Hee
and Kim, Timur K.
and Hoesch, Moritz
and Kim, Keun Su},
title={Holstein polaron in a valley-degenerate two-dimensional semiconductor},
journal={Nat. Mater.},
year={2018},
month={08},
day={01},
volume={17},
number={8},
pages={676-680},
issn={1476-4660},
doi={10.1038/s41563-018-0092-7},
url={https://doi.org/10.1038/s41563-018-0092-7}
}

@article{Leicht16,
author = {Leicht, Philipp and Zielke, Lukas and Bouvron, Samuel and Moroni, Riko and Voloshina, Elena and Hammerschmidt, Lukas and Dedkov, Yuriy S. and Fonin, Mikhail},
title = {In Situ Fabrication Of Quasi-Free-Standing Epitaxial Graphene Nanoflakes On Gold},
journal = {ACS Nano},
volume = {8},
number = {4},
pages = {3735-3742},
year = {2014},
doi = {10.1021/nn500396c},
URL = {https://doi.org/10.1021/nn500396c},

}

@article{Smith74,
  title = {Photoemission spectra and band structures of $d$-band metals. III. Model band calculations on Rh, Pd, Ag, Ir, Pt, and Au},
  author = {Smith, Neville V.},
  journal = {Phys. Rev. B},
  volume = {9},
  issue = {4},
  pages = {1365--1376},
  numpages = {0},
  year = {1974},
  month = {02},
  publisher = {American Physical Society},
  doi = {10.1103/PhysRevB.9.1365},
  url = {https://link.aps.org/doi/10.1103/PhysRevB.9.1365}
}

@article{safeer25b,
author = {Safeer, Affan and Güleryüz, Oktay and Atodiresei, Nicolae and Jolie, Wouter and Michely, Thomas and Fischer, Jeison},
title = {MnBr$_2$ on the Graphene on Ir(110) Substrate: Growth, Structure, and Super-Moiré},
journal = {Small},
pages = {e09471},
doi = {https://doi.org/10.1002/smll.202509471},
url = {https://onlinelibrary.wiley.com/doi/abs/10.1002/smll.202509471},
year = {2026}
}

@article{Elmaslmane18,
author = {Elmaslmane, A. R. and Watkins, M. B. and McKenna, K. P.},
title = {First-Principles Modeling of Polaron Formation in TiO$_2$ Polymorphs},
journal = {Journal of Chemical Theory and Computation},
volume = {14},
number = {7},
pages = {3740-3751},
year = {2018},
doi = {10.1021/acs.jctc.8b00199},
URL = {https://doi.org/10.1021/acs.jctc.8b00199},
eprint = {https://doi.org/10.1021/acs.jctc.8b00199}
}

@article{Bagchi2024,
  title = {Spin-Polarized Scanning Tunneling Microscopy Measurements of an Anderson Impurity},
  author = {Bagchi, Mahasweta and Tounsi, Tfyeche Y. and Safeer, Affan and van Efferen, Camiel and Rosch, Achim and Michely, Thomas and Jolie, Wouter and Costi, Theo A. and Fischer, Jeison},
  journal = {Phys. Rev. Lett.},
  volume = {133},
  issue = {24},
  pages = {246701},
  numpages = {6},
  year = {2024},
  month = {Dec},
  publisher = {American Physical Society},
  doi = {10.1103/PhysRevLett.133.246701},
  url = {https://link.aps.org/doi/10.1103/PhysRevLett.133.246701}
}

@article{Knispel2025,
  title = {Atomic-scale Dzyaloshinskii-Moriya-modified Yoshimori spirals in an Fe double layer on Ir(110)},
  author = {Knispel, Timo and Tseplyaev, Vasily and Bihlmayer, Gustav and Bl\"ugel, Stefan and Michely, Thomas and Fischer, Jeison},
  journal = {Phys. Rev. B},
  volume = {111},
  issue = {2},
  pages = {L020405},
  numpages = {6},
  year = {2025},
  month = {Jan},
  publisher = {American Physical Society},
  doi = {10.1103/PhysRevB.111.L020405},
  url = {https://link.aps.org/doi/10.1103/PhysRevB.111.L020405}
}

@article{Moue22 ,
author = {Moue, Rena and Kokubo, Yuta and Mukai, Kozo and Mizushima, Hirotaka and Fukushima, Yuto and Kawaguchi, Kaishu and Kondo, Takeshi and Harasawa, Ayumi and Yoshinobu, Jun and Mase, Kazuhiko and Shin, Shik and Kanai, Kaname},
title = {Enhanced Rashba Splitting of Au(111) Surface States with Hydrogen-Bonded Melamine-Based Organic Framework},
journal = {Advanced Materials Interfaces},
volume = {9},
number = {25},
pages = {2201102},
doi = {https://doi.org/10.1002/admi.202201102},
url = {https://advanced.onlinelibrary.wiley.com/doi/abs/10.1002/admi.202201102},
eprint = {https://advanced.onlinelibrary.wiley.com/doi/pdf/10.1002/admi.202201102},
year = {2022}
}

@article{Reinert01,
  title = {Direct measurements of the L-gap surface states on the (111) face of noble metals by photoelectron spectroscopy},
  author = {Reinert, F. and Nicolay, G. and Schmidt, S. and Ehm, D. and H{\"u}fner, S.},
  journal = {Phys. Rev. B},
  volume = {63},
  issue = {11},
  pages = {115415},
  numpages = {7},
  year = {2001},
  month = {Mar},
  publisher = {American Physical Society},
  doi = {10.1103/PhysRevB.63.115415},
  url = {https://link.aps.org/doi/10.1103/PhysRevB.63.115415}
}

@article{Tersoff83,
  title = {Theory and Application for the Scanning Tunneling Microscope},
  author = {Tersoff, J. and Hamann, D. R.},
  journal = {Phys. Rev. Lett.},
  volume = {50},
  issue = {25},
  pages = {1998--2001},
  numpages = {0},
  year = {1983},
  month = {Jun},
  publisher = {American Physical Society},
  doi = {10.1103/PhysRevLett.50.1998},
  url = {https://link.aps.org/doi/10.1103/PhysRevLett.50.1998}
}

@article{Emin73,
author = {David Emin},
title = {On the existence of free and self-trapped carriers in insulators: an abrupt temperature-dependent conductivity transition},
journal = {Advances in Physics},
volume = {22},
number = {1},
pages = {57--116},
year = {1973},
publisher = {Taylor \& Francis},
doi = {10.1080/00018737300101269},
URL = {https://doi.org/10.1080/00018737300101269},
eprint = {https://doi.org/10.1080/00018737300101269}
}

@article{Safeer2026Cr2S3,
author = {Safeer, Affan and Dias, Calisa and Ghorbani-Asl, Mahdi and Karaka, Abdallah and Bawankule, Pradyumna and Li, Weibin and Gargiani, Pierluigi and Jolie, Wouter and Krasheninnikov, Arkady V. and Bedoya-Pinto, Amilcar and Michely, Thomas and Fischer, Jeison},
title = {Atomic-Scale Detection of Néel Vector Switching in the Single-Layer A-Type Antiferromagnet Cr2S3-2D},
journal = {Advanced Materials},
volume = {38},
number = {58},
pages = {e74637},
doi = {https://doi.org/10.1002/adma.74637},
url = {https://advanced.onlinelibrary.wiley.com/doi/abs/10.1002/adma.74637},
year = {2026}
}




\end{document}


\DeclareGraphicsExtensions{.pdf}
\title{Supplemental Material:\\
Substrate manipulation of small polarons in single-layer MnBr$_2$}

\author{Affan Safeer}
\email{safeer@ph2.uni-koeln.de}
\affiliation{II. Physikalisches Institut, Universit\"{a}t zu K\"{o}ln, Z\"{u}lpicher Stra\ss e 77, 50937 K\"{o}ln, Germany}
\author{Oktay G\"{u}lery\"{u}z} 
\affiliation{II. Physikalisches Institut, Universit\"{a}t zu K\"{o}ln, Z\"{u}lpicher Stra\ss e 77, 50937 K\"{o}ln, Germany}
\author{Guangyao Miao} 
\affiliation{II. Physikalisches Institut, Universit\"{a}t zu K\"{o}ln, Z\"{u}lpicher Stra\ss e 77, 50937 K\"{o}ln, Germany}
\author{Wouter Jolie}
\affiliation{II. Physikalisches Institut, Universit\"{a}t zu K\"{o}ln, Z\"{u}lpicher Stra\ss e 77, 50937 K\"{o}ln, Germany}
\author{Keith McKenna}
\affiliation{School of Physics, Engineering and Technology, University of York, Heslington
York YO10 5DD, U.K.}
\author{Thomas Michely} 
\affiliation{II. Physikalisches Institut, Universit\"{a}t zu K\"{o}ln, Z\"{u}lpicher Stra\ss e 77, 50937 K\"{o}ln, Germany}
\author{Jeison Fischer}
\affiliation{II. Physikalisches Institut, Universit\"{a}t zu K\"{o}ln, Z\"{u}lpicher Stra\ss e 77, 50937 K\"{o}ln, Germany}


\maketitle
\newpage
\tableofcontents
\newpage

\maketitle
\newpage
\subsection*{Supplementary note 1: Extended dataset for bias-dependent polaron imaging and mobility.}
\begin{figure}[hbt!]
\includegraphics[width=\textwidth]{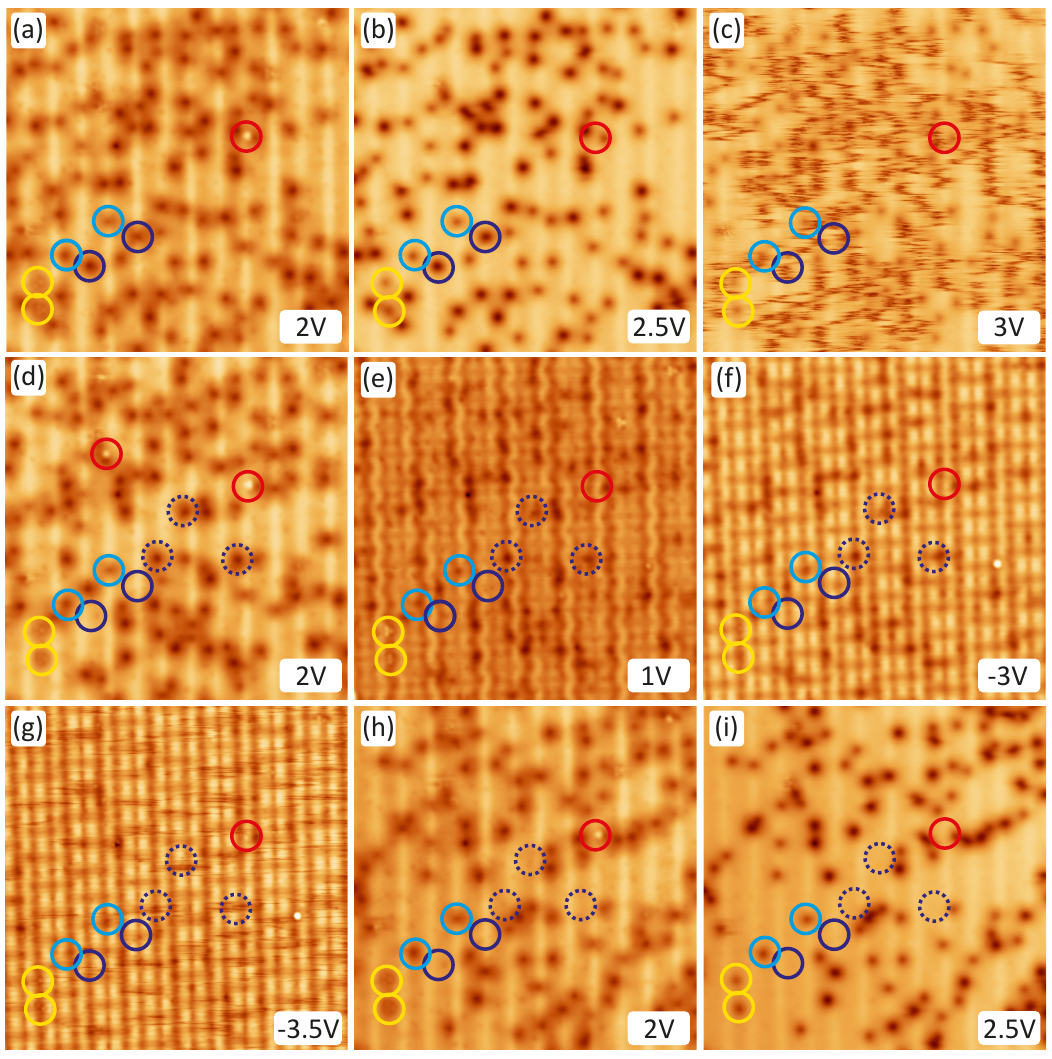}
\caption{(a-i) Sequence of STM images of the same area on MnBr$_2$/Gr/Ir(110) acquired at varying $V_\mathrm{b}$ ranging from $+3$\,V and $-3.5$\,V, as indicated, and at $I_\mathrm{t} = 50$\,pA. Selected polarons are encircled to track their position: dark blue (type-I mobile), light blue (type-I immobile), yellow (type-II 2x2), and red (type-II 1x1).
From comparison of (d) and (h) it is obvious that type-I mobile polarons are also moved at $V_\mathrm{b}= -3.5$\,V. Image sizes: 45\,nm $\times$ 45\,nm.}
\label{S3} 
\end{figure}

Figure~\ref{S3} presents a complementary dataset to Figure 2 of the main text, extending the investigation to a higher negative bias range. The behavior at positive $V_\mathrm{b}$ [Figure ~\ref{S3}(a-e)] is consistent with our previous observations. The four types of polarons are distinguishable at $V_\mathrm{b}$ = $2$\,V, and type-I mobile polarons change positions after increasing $V_\mathrm{b}$ to $3$\,V, as tracked by the polarons encircled dark blue in Figure ~\ref{S3}(a-d).

The tip-induced mobility is also observed at negative $V_\mathrm{b}$. If $V_\mathrm{b}$ is set to $-3.5$\,V in Figure~S1(g), type-I mobile polarons are mobile as obvious from the comparison of Figures~S1(d) and (h) (check position in dotted dark blue circles).

\maketitle
\newpage
\subsection*{Supplementary note 2: Mobility of type-I mobile polaron for lower positive $V_\mathrm{b}$ at large tunneling current.}
\begin{figure}[hbt!]
\includegraphics[width=\textwidth]{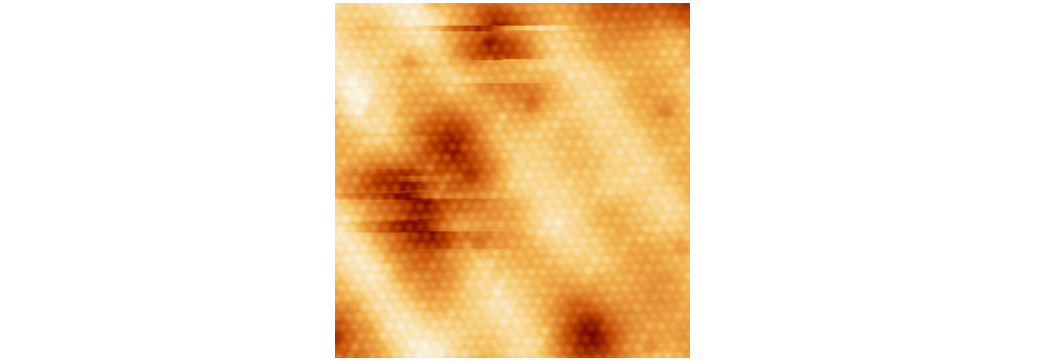}
\caption{STM image of MnBr$_2$/Gr/Ir(110) acquired with $V_\mathrm{b} = 2$\,V, $I_\mathrm{t} = 3$\,nA. Image size: 15\,nm $\times$ 15\,nm.}

\label{S2} 
\end{figure}

\newpage
\subsection*{Supplementary Note 3: Overview topographs of single-layer MnBr$_2$ grown on different substrates.}
\begin{figure}[hbt!]
\includegraphics[width=\textwidth]{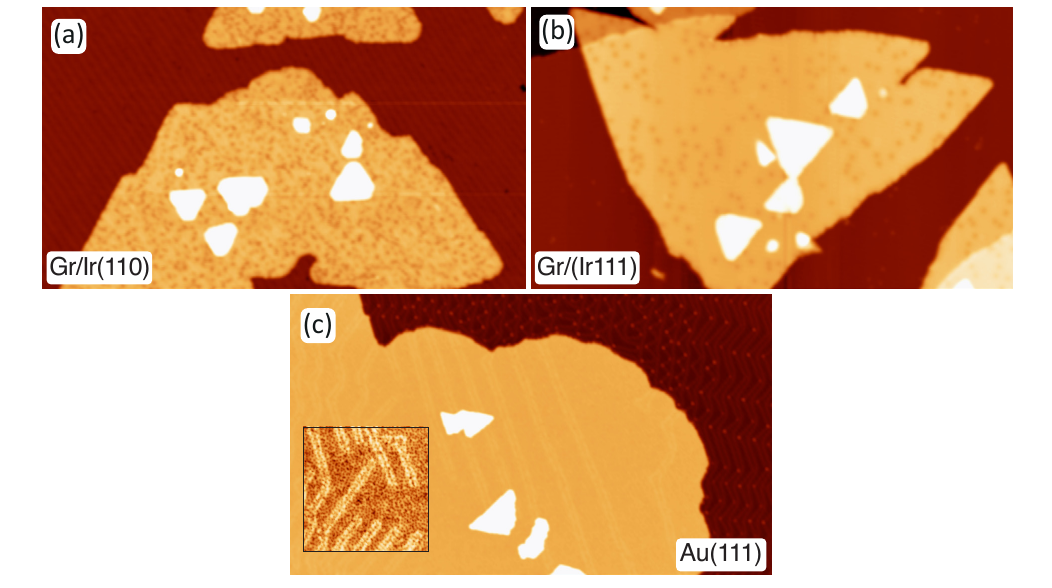}
\caption{\normalsize (a) Overview STM topographs of MnBr$_2$ on (a) Gr/Ir(110), (b) Gr/Ir(111) and (c) Au (111). A high-resolution zoomed STM image is pasted as an inset to (c) at the exact location it has in the overview. In the inset, polarons and the herringbone reconstruction are well visible. STM Imaging Parameters: (a) $V_\mathrm{b} = 3.5$\,V, $I_\mathrm{t} = 20$\,pA, (b) $V_\mathrm{b} = 3.5$\,V, $I_\mathrm{t} = 20$\,pA, (c) and inset of (c) $V_\mathrm{b} = 2$\,V, $I_\mathrm{t} = 50$\,pA. Image sizes: (a-c) 240\,nm $\times$ 140\,nm, inset in (c) 60\,nm $\times$ 60\,nm. }
  \label{fgr:SI_Fig_polarons_compare} 
\end{figure}

On top of single-layer MnBr$_2$ islands, small second-layer MnBr$_2$ islands are also visible. The overview STM image of MnBr$_2$ on Au(111) in Figure~S3(c) shows the Au(111)-herringbone reconstruction through the island, which is only partly lifted by MnBr$_2$ overgrowth.

\maketitle
\newpage
\subsection*{Supplementary note 4: Tip induced conversions of type-II 1x1 polarons into type-I mobile polarons.}
\begin{figure}[hbt!]
\includegraphics[width=\textwidth]{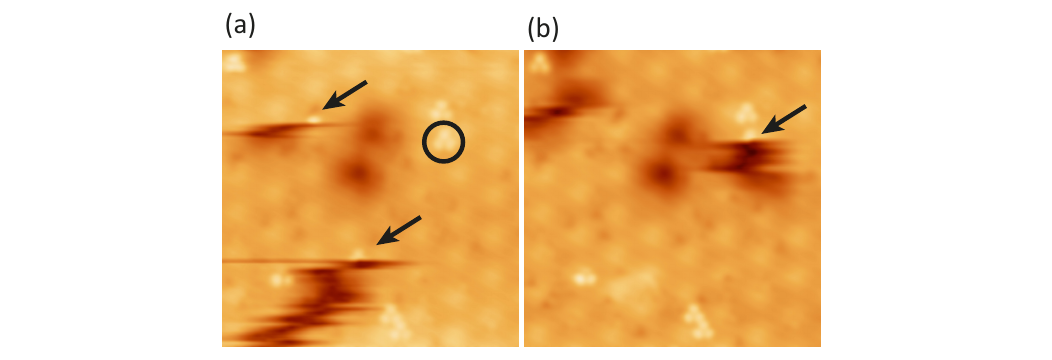}
\caption{(a,b) Consecutive STM topographs of MnBr$_2$ on Gr/Ir(111). The black arrows in (a) indicate locations where the STM tip induces a conversion from a type-II 1x1 polaron to a type-I mobile polaron. Additionally, the type-II 1x1 polaron circled in (a) has converted to a type-I mobile polaron in (b), as highlighted by the arrow in (b). STM imaging parameters: $V_\mathrm{b} = 2.5$\,V, $I_\mathrm{t} = 50$\,pA. Image sizes: 18\,nm $\times$ 18\,nm.}

\label{S4} 
\end{figure}

Figure~\ref{S4} demonstrates the conversion of type-II 1x1 polarons to type-I mobile polarons induced during STM scanning. After these conversions, the type-I mobile polarons move with the tip, appearing as fuzzy and continuous dark features in the topographs.

\maketitle
\newpage
\subsection*{Supplementary note 5: Creation of a type-I mobile polaron in MnBr$_2$ on Gr/Ir(111).}
\begin{figure}[hbt!]
\includegraphics[width=\textwidth]{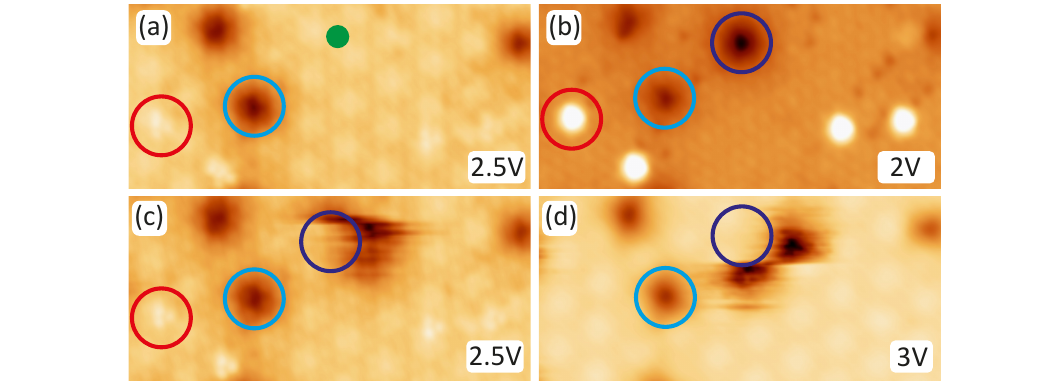}
\caption{Sequence of STM images illustrating the creation of a type-I mobile polaron in MnBr$_2$ on Gr/Ir(111) via bias pulsing. (a) STM topograph before bias pulsing with pre-existing type-II 1x1 and type-I immobile polarons (circled red and light blue, respectively). (b) STM topograph after bias pulse with $V_\mathrm{b} = +3.5$\,V and $I_\mathrm{t} = 500$\,pA at green dot in (a). (c,d) Subsequent imaging at higher biases shows the mobile polaron moving under the influence of the STM tip.  $I_\mathrm{t} = 50$\,pA. Image sizes: 22\,nm $\times$ 10\,nm.}

\label{S5} 
\end{figure}

Figure~\ref{S5} shows the creation of a type-I mobile polaron. A bias pulse ($V_\mathrm{b} = +3.5$\,V, $I_\mathrm{t} = 500$\,pA) is applied at the green spot in Figure~\ref{S5} (a). Subsequent imaging at $V_\mathrm{b} = +2$\,V, as shown in Figure~\ref{S5}(b), reveals a type-I mobile polaron (encircled purple) at the pulse location along with pre-existing type-I immobile (encircled blue) and type-II 1x1 polarons (encircled red). In subsequent imaging at $V_\mathrm{b} = +2.5$\,V and $+3\,V$ in Figure~\ref{S5}(c) and (d) the observed mobility of the polaron justifies its classification as type-I mobile polaron. 

\newpage

\maketitle
\newpage
\subsection*{Supplementary note 6: Polarons in MnBr$_2$ on Gr/Ir(111).}
\begin{figure}[hbt!]
\includegraphics[width=\textwidth]{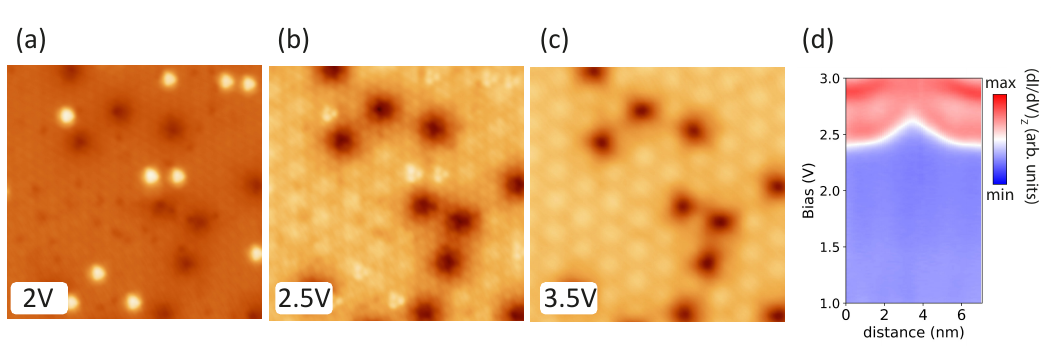}
\caption{(a-c) Consecutive STM topographs of MnBr$_2$ on Gr/Ir(111) at different $V_\mathrm{b}$ as indicated. (d) Constant current $\mathrm{d}I/\mathrm{d}V$ linescan map over the type-I immobile,  Spectra are acquired with $V_{st}$ = +3 V, $I_{st}$ = 50\,pA, $f_{mod}$ = 667\,Hz, and $V_{mod}$ = 20\,mV. STM imaging parameters: $I_\mathrm{t} = 50$\,pA. Image sizes: 18\,nm $\times$ 18\,nm.}

\label{S6} 
\end{figure}

\maketitle
\newpage
\subsection*{Supplementary note 7: Point defects in MnBr$_2$.}
\begin{figure}[hbt!]
\includegraphics[width=\textwidth]{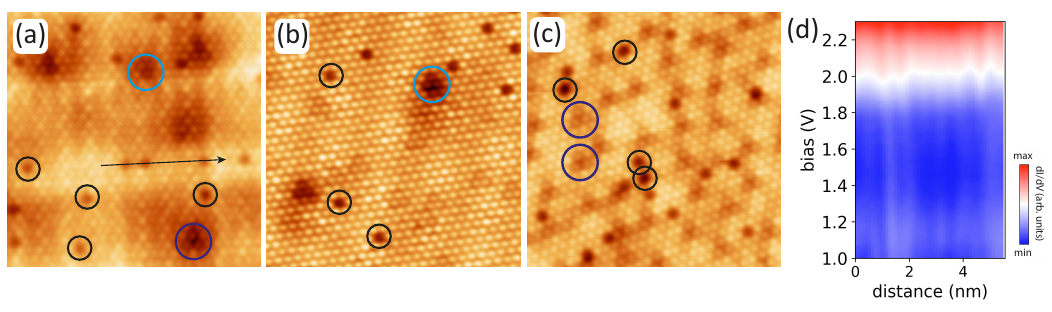}
\caption{Atomically resolved STM topographs MnBr$_2$ grown on (a) Gr/Ir(110), (b) Gr/Ir(111), and (c) Au(111). (d) Constant current  $\mathrm{d}I/\mathrm{d}V$ linescan over the point defects along the arrow in (a). Spectra are acquired with $V_{st}$ = +2.3 V, $I_{st}$ = 50\,pA, $f_{mod}$ = 667\,Hz, and $V_{mod}$ = 20\,mV. STM imaging parameters: (a) $V_\mathrm{b} = 2$\,V and $I_\mathrm{t} = 100$\,pA, (b) $V_\mathrm{b} = -2$\,V and $I_\mathrm{t} = 50$\,pA and (c) $V_\mathrm{b} = 2.5$\,V and $I_\mathrm{t} = 50$\,pA  and STM images size: 10\,nm $\times$ 10\,nm.  
} 
\label{Sfig_point_defect} 
\end{figure}

Figure~\ref{Sfig_point_defect} shows atomically resolved STM images of MnBr$_2$ grown on Gr/Ir(110), Gr/Ir(111), and Au(111), where point defects (some encircled in black) are clearly visible and distinct from polarons (encircled in light and dark blue). When the $\mathrm{d}I/\mathrm{d}V$ linescan is taken over these point defects, no upward band bending is observed, as shown in Figure~\ref{Sfig_point_defect}(d).

\maketitle
\newpage
\subsection*{Supplementary note 8: Spatial localization of polaron in MnBr$_2$.}
\begin{figure}[hbt!]
\includegraphics[width=\textwidth]{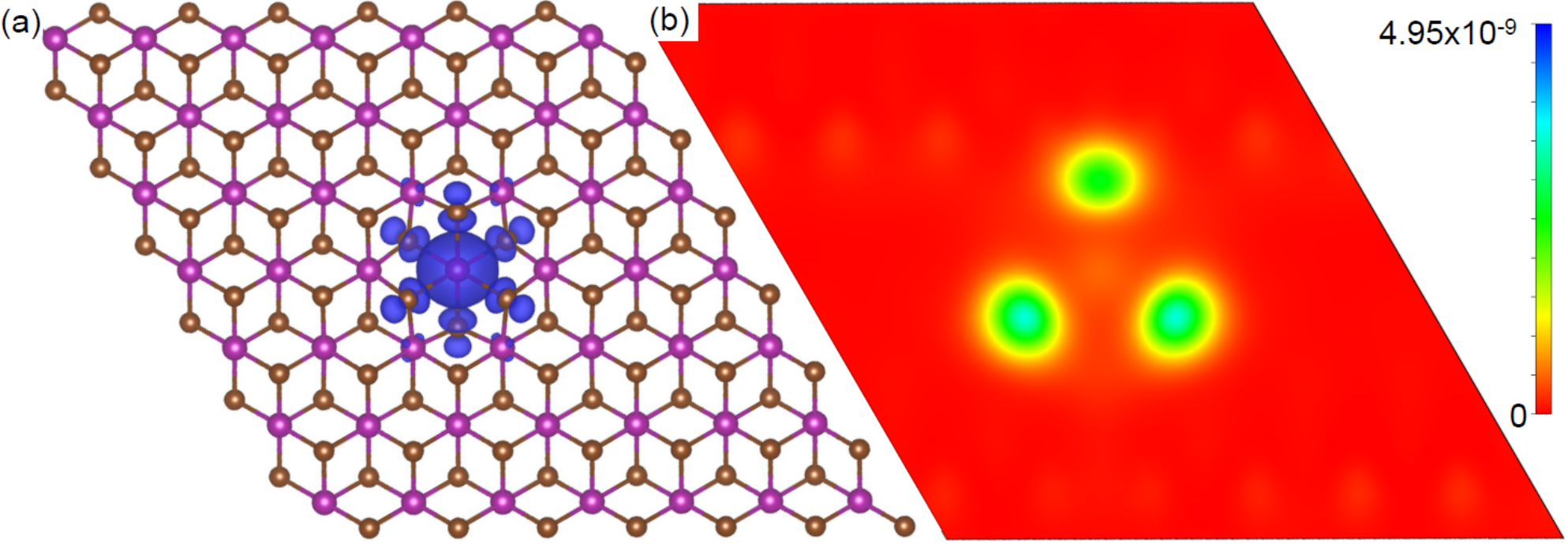}
\caption{(a) Electron spin-density isosurface showing the spatial localization of the small electron polaron in MnBr$_2$. Mn atoms are shown as pink spheres and Br atoms as brown spheres.
(b) Simulated STM topograph of the small polaron in (a). Cross section of the electron density associated with the in-gap polaron state at a height of 3\,\AA~above the center of the MnBr$_2$ layer (color scale units of electrons per \AA$^2$). This image provides a simple prediction of the expected STM contrast within the Tersoff-Hamann approximation \cite{Tersoff83}, assuming tunneling is dominated by the overlap between the wavefunction of the tip and the in-gap polaron state.
} 
\label{SI-polarom-DFT} 
\end{figure}

Details of density functional theory calculations:\\
Density functional theory calculations were carried out using the projector augmented-wave (PAW) method, a plane-wave basis with a cut-off of 520 eV as implemented in the Vienna Ab Initio Simulation Package ({\sc vasp} 6.5.1) \cite{kresse_efficiency_1996, kresse_efficient_1996}. We employed the Heyd-Scuseria-Ernzerhof hybrid functional\cite{HSE}\ which gives an improved description of band gaps and reduced self-interaction error compared to semi-local exchange-correlation functionals. The bulk structure of the MnBr$_2$ monolayer (in an antiferromagnetic configuration) was optimized using a $2\times2$ supercell, a vacuum gap of 20 \AA , and a $3\times3\times1$ k-point grid, yielding a lattice constant of 3.88 \AA . To model polaron formation, we employed a $6\times6$ supercell, added an additional electron, and performed precursor atomic distortions around a Mn site before full optimization. We were unable to find a localized small polaron using the default 25\%\ Hartree-Fock (HF) exchange or with 30\% HF exchange. However, for 35\%\ HF exchange and above a local energy minimum corresponding to a self-trapped small polaron is found. The trapping energy with respect to a delocalized electron in the conduction band for 35\%\ HF exchange is +0.18 eV (i.e., it is metastable). The predicted band gap of the MnBr$_2$ monolayer also increases with increasing HF exchange: 4.5 eV (25\% ), 5.0 eV (30\% ), 5.4 eV (35\% ), and 5.9 eV (40\% ). The {\sc sumo} Python package was used for density of states plots \cite{ganose_sumo_2018}\, and the atomic structures and spin-density iso-surfaces were visualized using the {\sc vesta} software package \cite{momma_vesta_2008}.

\maketitle
\newpage
\subsection*{Supplementary note 9: Electronic density of states for polaron in MnBr$_2$.}
\begin{figure}[hbt!]
\includegraphics[width=0.6\textwidth]{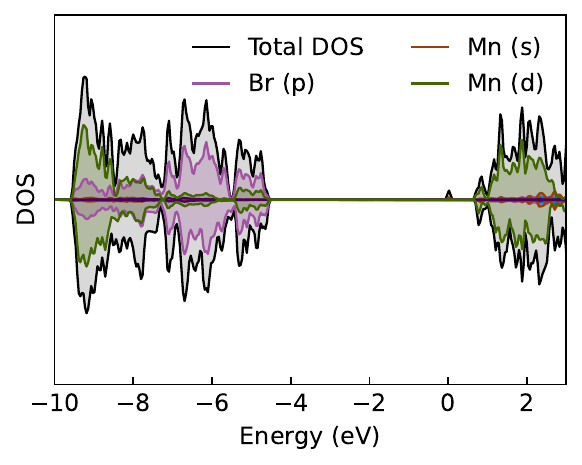}
\caption{Electronic density of states calculated for the small electron polaron configuration in MnBr$_2$ (obtained using the HSE hybrid functional and 35\%\ Hartree-Fock exchange).
} 
\label{SI-Polaron-DOS-DFT} 
\end{figure}

\maketitle
\newpage
\subsection*{Supplementary note 10: Bias-dependent absolute tip heights above Gr/Ir(110) and MnBr$_2$/Gr/Ir(110).}
\begin{figure}[hbt!]
\includegraphics[width=\textwidth]{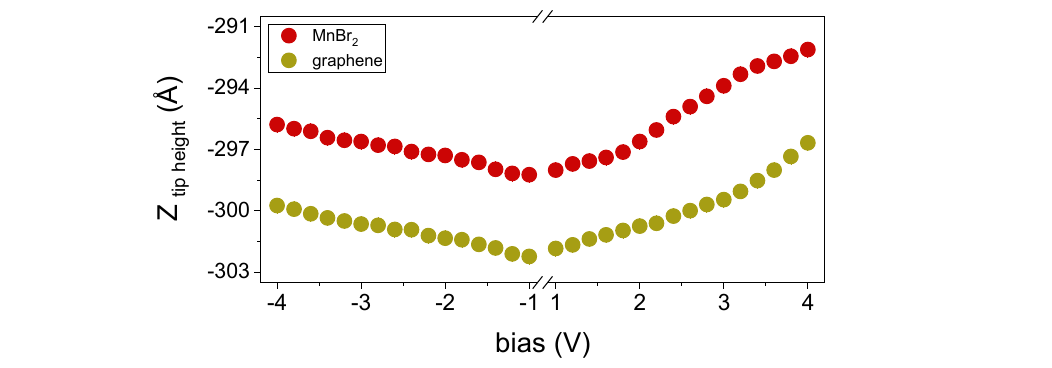}
\caption{Absolute heights of the STM tip over Gr/Ir(110) (red squares, right $y$-axis) and MnBr$_2$/Gr/Ir(110) (black squares, left $y$-axis) for $I_\mathrm{t} = 20$\,pA as a function of $V_\mathrm{b}$.
} 
\label{Sfig_heights} 
\end{figure}

\newpage

\bibliography{Ref}